\documentclass[%
reprint,
superscriptaddress,
 amsmath,amssymb,
prr,
]{revtex4-2}
\usepackage{graphicx}
\usepackage{dcolumn}
\usepackage{bm}
\usepackage{url}

\usepackage{color}
\usepackage[normalem]{ulem} 
\usepackage{multirow}
\usepackage{placeins}

\makeatletter
\def\maketitle{
\@author@finish
\title@column\titleblock@produce
\suppressfloats[t]}
\makeatother

\begin{document}
\preprint{APS/123-QED}

\title{Order-$N$ orbital-free density-functional calculations with machine learning of functional derivatives for semiconductors and metals}

\author{Fumihiro Imoto$^{1,2}$, Masatoshi Imada$^{2,3}$ and Atsushi Oshiyama}
\affiliation{Institute of Materials and Systems for Sustainability, Nagoya University, Nagoya 464-8603, Japan \\
$^2$Waseda Research Institute for Science and Engineering, Waseda University, Tokyo 169-8555, Japan \\
$^3$Toyota Physical and Chemical Research Institute, Nagakute, 480-1192, Japan}
\date{\today}

\begin{abstract}
Orbital-free density functional theory (OFDFT) offers a challenging way of electronic-structure calculations scaled as $\mathcal{O}(N)$ computation for system size $N$. We here develop a scheme of the OFDFT calculations based on the accurate and transferrable kinetic-energy density functional (KEDF) which is created in an unprecedented way using appropriately constructed neural network (NN). We show that our OFDFT scheme reproduces the electron density obtained in the state-of-the-art DFT calculations and then provides accurate structural properties of 24 different systems, ranging from atoms, molecules, metals, semiconductors and an ionic material. The accuracy and the transferability of our KEDF is achieved by our NN training system in which the kinetic-energy functional derivative (KEFD) at each real-space grid point is used. The choice of the KEFD as a set of training data is essentially important, because first it appears directly in the Euler equation which one should solve and second, its learning assists in reproducing the physical quantity expressed as the first derivative of the total energy. More generally, the present development of KEDF $T[\rho]$ is in the line of systematic expansion in terms of the functional derivatives $\delta^{\ell_1} T/\delta \rho^{\ell_1}$ through progressive increase of $\ell_1$. The present numerical success demonstrates the validity of this approach. The computational cost of the present OFDFT scheme indeed shows the $\mathcal{O}(N)$ scaling, as is evidenced by the computations of the semiconductor SiC used in power electronics. 
\end{abstract}
\maketitle

\section{Introduction}\label{sec:intro}

Density Functional Theory (DFT) \cite{hohenberg-kohn-pr64} proves that the ground-state total energy $E$ of an interacting electron system is a unique universal functional $G [\rho]$ of the electron density $\rho (\bm r)$ plus the electrostatic energy $V_{\rm ext} [\rho]$ under the external potential $v_{\rm ext} (\bm r)$, opening a possibility to compute physical properties of real materials by solving an Euler equation $\delta E [\rho] / \delta \rho (\bm r) = \mu$, where $\mu$ is the Lagrange multiplier that enforces density normalization. Various attempts have been made by introducing virtual non-interacting electron systems, in which the electron densities are 
identical to those in corresponding real materials, 
and then decomposing $G$ to the kinetic energy of the noninteracting system $T_s [\rho]$, the classical electron-electron interaction energy $E_{\rm H} [\rho]$, and the remaining exchange-correlation energy $E_{\rm xc} [\rho]$ \cite{kohn-sham-pr65}. In this scheme, $T_s$ is expressed as a sum of the kinetic-energy contribution from each orbital $\phi_i (\bm r)$ (Kohn-Sham orbital) as, 
\begin{equation}\label{T_ks}
T_s [\rho] = \frac{1}{2} \sum_{i} \int | \nabla \phi_{i}({\bm r}) |^{2} d{\bm r}, 
\end{equation}
and thus the original Euler equation in DFT, 
\begin{equation}\label{Euler-ofdft}
 \frac{\delta T_s [\rho]}{\delta \rho({\bm r})} + v_{\rm{ext}}({\bm r}) 
+ \frac{\delta E_{\rm{H}}[\rho]}{\delta \rho({\bm r})} 
+ \frac{\delta E_{\rm{xc}}[\rho]}{\delta \rho({\bm r})} = \mu,
\end{equation}
becomes a set of Schr\"{o}dinger-like equations (Kohn-Sham equations) which in turn determine $\phi_i$ self-consistently.

A numerous number of works adopting this Kohn-Sham (KS) scheme (KSDFT) has been applied to a various materials and achieved unprecedented 
success \cite{jones2015,mardirossian-molphys17}, 
depending on the level of the approximation to the exchange-correlation functional (Jacob's ladder) \cite{JacobLadder}. However,  solving the Kohn-Sham equations for all the occupied orbitals in the system is a computational burden scaling with the system size $N$ as $\mathcal{O}(N^3)$, thus restricting the applicability of DFT. The scheme with lower-order scaling is highly demanded in materials science and also in advancing DFT. One of the solutions in a legitimate way is the orbital-free density-functional theory (OFDFT) in which $T_s [\rho]$ is expressed as a functional of $\rho$, the kinetic energy density functional (KEDF), and the Euler equation Eq.~(\ref{Euler-ofdft}) 
remains as a single equation for $\rho$, thus OFDFT being expected to be $\mathcal{O}(N)$ scheme. 

Such OFDFT approach free from the orbitals is in the heart of the original DFT \cite{hohenberg-kohn-pr64} and was initiated much before DFT, being known as Thomas-Fermi (TF) theory for the homogeneous electron gas \cite{thomas-proccamphilsoc27,fermi-zphys28} and von Weizs\"{a}cker (vW) gradient expression \cite{vweizsacker-zphys35}. 
Based on these approaches, the kinetic energy functional $T_s$ is generally written as 
\begin{equation}\label{def_enhance_fac}
 T_s [\rho] = \int \tau^{\rm TF} ({\bm r}) F [\rho] d{\bm r}, 
\end{equation}
where $\tau^{\rm TF} ({\bm r})$ is the kinetic energy density in the TF approximation, $ \tau^{\rm TF} = (3/10)(3\pi^{2})^{2/3} \rho^{5/3}$, and $F$ is so called the enhancement factor. 

There have been a lot of efforts to develop the enhancement factor \cite{schwartz02,bach-dellesite14} either in a semilocal \cite{semilocalKEDF,constantin-PGSL-jpclett18,constantin-PGSL-jctc19,luo-LKT-prb18} or a nonlocal form \cite{wang-teter-prb92, garcia-aldea-alvarellos-pra07, gargonz-alvarellos-chacon-prb96, wang-govind-carter-prb98, wang-govind-carter-prb99, zhou-ligneres-carter-jchemphys05, garcia-aldea-alvarellos-pra07, garcia-aldea-alvarellos-pra08, huang-carter-prb10} to reproduce the KS kinetic energy $T^{\rm KS}$ as accurate as possible. 
Among the various semilocal forms, recent two functionals, PGSL$\beta$ (Pauli-Gaussian second order and Laplacian with a parameter $\beta$) \cite{constantin-PGSL-jpclett18,constantin-PGSL-jctc19} and LKT (Luo-Karasiev-Trickey) \cite{luo-LKT-prb18}, reproduce experimental structural properties successfully within the error of less than a few percent for the lattice constants and of about ten percent for the bulk moduli. 
Semilocal KEDFs have been usually developed in the form of the generalized gradient approximation (GGA) and satisfy some of the exact conditions for
(a) the small limit of the density gradient in the gradient expansion (GE), (b) the large limit of the density-gradient in GE, (c) the positivity of Pauli potential \cite{levy-ouyang-pra88}, and (d) the linear response function in the homogeneous density limit, namely being equal to  the Lindhard function \cite{lindhard}.  
PGSL$\beta$ satisfies the first three conditions (a)--(c). The parameter $\beta$ is fitted so as to minimize the mean absolute relative errors of physical quantities such as the cell volume, the bulk modulus, the total energy at the equilibrium volume, and the electron density, with respect to 
the corresponding values obtained in the KS scheme, leading to the value of $\beta=0.25$ \cite{constantin-PGSL-jpclett18,constantin-PGSL-jctc19}. 
LKT satisfies the conditions (b) and (c), and is left with a single parameter ``$a$''.
The allowable range of ``$a$'' is determined by the condition (c) for atomic densities generated by pseudopotentials. 
The obtained range is $0.0 \le a \le 1.3$ and the value of $a = 1.3$ is typically used \cite{luo-LKT-prb18}. 
Karasiev {\it et al.} \cite{karasiev-VT84F-prb13} proposed another semilocal parameter-free KEDF that satisfies the conditions (a)--(c). 
However, The bulk moduli from this functional are around 50 \% higher than the reference KS values. Hence, we will not include the results by this functional in our benchmark in the present paper.

Inclusion of the nonlocal effects in the enhancement factor indeed improves the performance of KEDF \cite{wang-govind-carter-prb99,wang-govind-carter-prb98,huang-carter-prb10,shin-carter-jcp140,mi-genova-pavanello-jcp18}. 
The nonlocal functional typically reproduces the structural properties within the error of less than 1 \% for the lattice constants and 10 \% for the bulk moduli \cite{mi-genova-pavanello-jcp18}. 
Nonlocal KEDFs such as WGC (Wang-Govind-Carter) \cite{wang-govind-carter-prb99} and related nonlocal KEDFs \cite{wang-govind-carter-prb98, wang-teter-prb92, garcia-aldea-alvarellos-pra07, gargonz-alvarellos-chacon-prb96, zhou-ligneres-carter-jchemphys05, garcia-aldea-alvarellos-pra07, garcia-aldea-alvarellos-pra08} were constructed to satisfy the conditions (a), (b), and (d). 
HC (Huang-Carter)\cite{huang-carter-prb10} functional improves the performance of WGC and its relatives in semiconductors by using two parameters adjusted to reproduce bulk moduli, equilibrium volumes, and equilibrium energies by KSDFT. 
EvW-WGC (enhanced von Weizs\"{a}cker Wang-Govind-Carter) \cite{shin-carter-jcp140} functional is an another extension of WGC, which is a linear combination of vW, TF and WGC, where a system-dependent parameter determines the portions of each KEDFs. EvW-WGC is more accurate than HC if an optimally adjusted parameter is used. 
MGP (Mi-Genova-Pavanello)\cite{mi-genova-pavanello-jcp18} functional uses a unique way of imposing the condition (d), namely the functional integration of the inverse of the Lindhard function in homogeneous density limit. In spite of the better performance generally observed, the nonlocal scheme requires the heavier computational cost, which scales as $\mathcal{O}(N\log{N})$
and, if possible, the true $\mathcal{O}(N)$ scheme enabled by the semilocal scheme is desired.

In this paper, to establish an alternative scheme that follows strictly the $\mathcal{O}(N)$ scaling, we propose a new enhancement factor in KEDF by neural-network (NN) machine learning. Without resorting to the reproducibility of the structural properties of target materials, we focus on reproducing the electron density which is the quantity of basics in DFT. We show that our NN which is trained only with the electron density in diamond generates a KEDF and successfully reproduces structural properties of a variety of materials, thus demonstrating the potential of the machine learning for further developments of the density functional. 

Coming back to the DFT itself, the energy as a functional of the electron density, KEDF in the present case, is primary. Its functional derivative defines the Euler equation and, on the other hand, provides the physical properties related to the first order derivative of the total energy. The higher derivatives are obviously important to describe the linear and nonlinear responses of materials. It is thus desirable to construct KEDF as well as its functional derivatives by term-by-term conformation of the kinetic-energy functional-derivative (KEFD) $\delta^{\ell_1} T/\delta \rho^{\ell_1}$ to the KS derivative through progressive increase of $\ell_1$. This is combined with the variational optimization of KEDF as a functional of both $\rho$ and its spatial derivatives $d^{\ell_2}\rho/d\bm{r}^{\ell_2}$ with increasing $\ell_2$. This systematic approach was formidable in the past but now may be practical using the machine learning. We here demonstrate a success of our first attempt with $\ell_1=1$ and $\ell_2=2$. 

The organization of the present paper is as follows. 
In \ref{sec:methods}, we explain the way of constructing our NN KEDFs. 
Computational details are also explained. 
The obtained NN-KEDFs are applied to 24 systems which include 7 atoms, 6 diatomic molecules and 11 solids ranging from metals, semiconductors to an ionic solid in \ref{sec:results}. 
The accuracy of our NN KEDFs is assessed in detail and discussed. 
In \ref{sec:results}, we also demonstrate the $\mathcal{O}(N)$ computational time scaling of our orbital-free implementation achieved for the system with thousands of atoms. 
We summarize our findings in \ref{sec:conc}.

\section{Methods}\label{sec:methods}

\subsection{Neural network for developing kinetic energy density functional}\label{ssec:nn_def}
\begin{figure}
\begin{center}
\includegraphics[width=.9\linewidth]{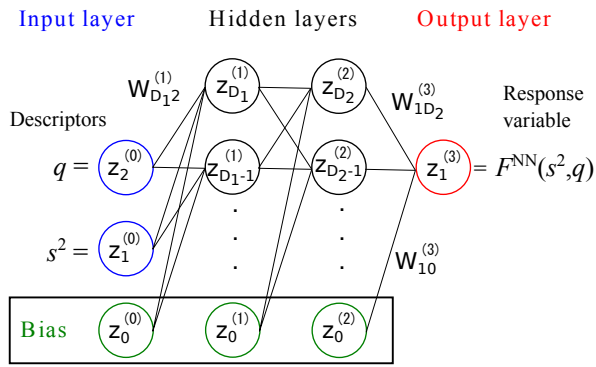}
\caption{
Schematic structure of the NN with two hidden layers ($N=3$).}\label{NN_structure}
\end{center}
\end{figure}

The neural network (NN) in general consists of an input layer, multiple hidden layers, and an output layer. Each of those layers with an index $l\;(l=0,\ldots,N)$ 
is composed of neurons with indices $k\;(k=0,\ldots,D_{l})$, where $N-1$ is the number of hidden layers and 
$D_{l}$ is the number of neurons in the $l$-th layer (Fig.~\ref{NN_structure}). 
The output from the $j$-th neuron in the $l$-th layer ($1\le l \le N-1$) 
is generally written as 
\begin{equation}\label{NN_general}
 z_{j}^{(l)}=\sigma^{(l)}(a_{j}^{(l)})=\sigma^{(l)}\left( \sum_{k=0}^{D_{l-1}} W_{jk}^{(l)} z_{k}^{(l-1)}\right), 
\end{equation}
where $\sigma^{(l)}$ is the activation function 
with the variable $a_{j}^{(l)}$ being $\sum_{k=0}^{D_{l-1}} W_{jk}^{(l)} z_{k}^{(l-1)}$. 
In our notation, the inputs are $z_{j}^{(0)}$ ($j=1,2$) and the output is $z_{1}^{(N)}$. 
The parameters trained in the NN, $W_{jk}^{(l)}\;(1\le l\le N,\;1\le j\le D_{l},\;0\le k\le D_{l-1})$, are called weights. 
For each layer, by using additional inputs $z_{0}^{(l)}\equiv1$, $W_{j0}^{(l)}$ works as the bias parameter.

In order to develop our NN KEDF, we seek for the enhancement factor $F^{\rm NN}$ as a functional of dimensionless quantities derived from the gradient and Laplacian of the electron density, i.e., 
$s^2$ and $q$, (namely, up to $\ell_2=2$) 
where $s = |\nabla\rho| / [2(3\pi^{2})^{1/3}\rho^{4/3}]$ and 
$q = \nabla^{2} \rho/[4(3\pi^{2})^{2/3}\rho^{5/3}]$. 
The KEDF with this enhancement factor, 
\begin{equation}\label{NN_enhance_fac}
 T^{\rm{NN}}[{\rho}]=\int \tau^{\rm{TF}}({\bm r}) F^{\rm{NN}}(s^{2}, q ; \bm W) d{\bm r}, 
\end{equation}
satisfies the uniform scaling condition \cite{levy-perdew-pra85}. 
Here, the generic notation $\bm W \equiv \{ W^i ; i=1 \ldots n_{W} \}$ represents the set of weight parameters in the NN, 
\{ $W_{jk}^{(l)}$ \} in Eq. (\ref{NN_general}). 
Figure~\ref{NN_structure} shows the schematic structure of the NN with two inputs $s^{2}$ and $q$. 
The output is the enhancement factor $F^{\rm NN}(s^{2},q)$ introduced in Eq.~(\ref{NN_enhance_fac}). 
The inputs and outputs are of course spatial dependent and thus the functions of the real-space position ${\bm r}$.

The explicit formula of the enhancement factor in the present NN KEDF is, e.g., for $D_l = D$ and $N = 4$ 
(see subsection \ref{ssec:nn_train} for determination of the structure of our present NN), 
\begin{eqnarray}\label{NN_KEDF_explicit}
 F^{\rm NN}(s^{2},q)&=&\sum_{i=0}^{D} W_{1i}^{(4)}\sigma \left( \sum_{j=0}^{D} W_{ij}^{(3)} \right. \nonumber \\
 && \times \sigma\left. \left( \sum_{k=0}^{D}W_{jk}^{(2)}\sigma\left( \sum_{l=0}^{2}W_{kl}^{(1)}z_{l}^{(0)} \right) \right) \right), 
\end{eqnarray}
where $(z_{1}^{(0)},z_{2}^{(0)})=(s^2,q)$. 
In the present paper we adopt the activation function defined as the exponential linear unit (ELU)\cite{ELU_activation}
\begin{equation}
 \sigma(a_{j}^{(l)})=
 \begin{cases} a_{j}^{(l)},  & \mbox{if }a_{j}^{(l)}>0 \\ \exp(a_{j}^{(l)})-1, & \mbox{if }a_{j}^{(l)}\le0  \end{cases} 
\end{equation}
for each layer ($1\le l\le N-1$) and an identity function for an output layer ($l=N$).

We adopt the enhancement factor $F^{\rm NN}(s^{2},q)$ as the output variable $z_{1}^{(N)}$. However, $F^{\rm NN}(s^{2},q)$ is not directly learned as the training data. Instead, the KEFD with $\ell_1=1$ from the KS scheme, $\delta T^{\rm KS}/\delta \rho$, is used as the training data since we adopt the cost function defined in terms of the KEFD [Eq.~(\ref{cost_function}) below] is used as a criterion of the performance of our NN in the present paper. 

\subsection{Density-functional computation scheme}\label{computation}

Actual Density-Functional (DFT)  computations in the construction of our NN and its validation have been performed by our real-space code RSDFT \cite{RSDFT1,RSDFT2,RSDFT3} which is highly optimized for the massively parallel architecture. In RSDFT scheme, all the quantities are computed on grid points in real space and the converged results are obtained by reducing the grid spacing systematically. Thanks to the real-space treatments, the scheme is essentially free from Fourier Transform, which releases heavy communication burden in the massively-parallel architecture of computers. For the exchange-correlation functional, we use generalized gradient approximation by Perdew, Burke, and Ernzerhof (PBE) \cite{perdew-burke-ernzerhof-prl96}. 

In usual first-principles DFT calculations, orbital-dependent nonlocal pseudopotentials (NLPSs) are used to simulate nuclei and core electrons in real materials \cite{hsc-prl79,bhs-82}. 
Yet in the OFDFT scheme, the orbitals are unavailable so that we need to construct {\it ab-initio local ionic} pseudopotentials (LIPSs) to simulate nuclei and core electrons. We have newly generated such LIPSs following the scheme by Carter and collaborators \cite{zhou-wang-carter-prb04,huang-carter-pccp08,huang-carter-prb10,delrio-dieterich-carter-jctc17}. Our scheme to generate LIPS consists of (i) the generation of the electron density of the target element in KSDFT scheme, (ii) obtaining the effective potential from the above density by solving the inverse problem with the Kadantsev-Stott method \cite{kadantsev-stott-pra04}, and (iii) proper skeletonizing of the effective potential (Appendix ~\ref{append:LIPS} for details). It should be emphasized that our construction scheme is totally free from adjustable parameters. We have generated LIPSs for 7 elements, Li, C, Na, Al, Si, Cl, and Cu. The validity and transferability of the obtained LIPSs are evidenced by the electron densities and the structural properties of 8 various materials, bcc-Li, fcc-Al, fcc-Cu, bcc-Na, NaCl, ds-Si, ds-C, and zincblende(3C)-SiC (Appendix \ref{append:LIPS}).

\subsection{Training of neural network toward kinetic energy functional derivative}\label{ssec:nn_train}

In order to develop the KEDF by our NN that best reproduces the KS-KEDF, $T^{\rm KS} [\rho]$, within the framework of $\ell_1=1$, we minimize during the training process the cost function for KEFD, 
\begin{equation}\label{cost_function}
 L = \frac{1}{N_t} \sum_{n=1}^{N_t}\frac{1}{2} 
 \left[\frac{\delta T^{\rm{NN}}({\bm r}_{n})}{\delta \rho} -\frac{\delta T^{\rm{KS}}({\bm r}_{n})}{\delta \rho} \right]^2, 
\end{equation}
which is the mean-squared error between the kinetic energy functional derivative (KEFD), $\delta T^{\rm NN}/\delta \rho$, obtained by our NN and KEFD, $\delta T^{\rm KS}/\delta \rho$, from the KS orbitals $\{\phi_{i}({\bm r})\}$. 
Here, $N_t$ is the total number of the training data at the real space position ${\bm r}_{n}$. The analytical expressions of KEFDs are given by (see Appendix \ref{append:KEFD_formula} for the derivation), 
\begin{eqnarray}\label{def_KEFD_NN}
 \frac{\delta T^{\text{NN}}}{\delta \rho}
 & = & c_{0} \rho^{2/3} \left[\frac{5}{3} F^{\text{NN}}-\frac{8}{3}s^{2} \frac{\partial F^{\text{NN}}}{\partial(s^{2})}
 - \frac{5}{3} q \frac{\partial F^{\text{NN}}}{\partial q} \right] \nonumber \\
 & - & \frac{3}{20}\nabla \cdot \left[ \frac{\partial F^{\text{NN}}}{\partial(s^{2})} \frac{\nabla\rho}{\rho} \right]
 + \frac{3}{40}\nabla^{2}\left( \frac{\partial F^{\text{NN}}}{\partial q} \right), 
\end{eqnarray}
with $c_{0}=3(3\pi^{2})^{2/3}/10$, and 
\begin{eqnarray}\label{def_KEFD_KS}
 \frac{\delta T^{\rm KS}[\rho^{\rm KS}]}{\delta \rho}
 =\frac{1}{\rho^{\text{KS}}} \sum_{i} f_{i} \left[ -\frac{1}{2}\phi^{*}_{i}\nabla^{2}\phi_{i} \right. \nonumber \\
 \left.+(\varepsilon^{\text{HOKS}}-\varepsilon_{i})|\phi_{i}({\bm r})|^{2} \right],
\end{eqnarray}
where $\varepsilon^{\text{HOKS}}$ is the highest-occupied KS eigenvalue, and 
$\varepsilon_{i}$ and $f_{i}$ are the KS eigenvalue and occupation number 
of the $i$-th KS orbital, and $\rho^{\text{KS}}=\sum_{i}f_{i}|\phi_{i}|^{2}$, respectively \cite{levy-ouyang-pra88}.
Instead of directly learning $F^{\rm NN}(s^{2},q)$ from the training data, we minimize the cost function Eq.~(\ref{cost_function})
between KEFDs because we observe that the functional derivative is poorly reproduced when KEDF is trained\cite{imoto-phd-thesis19} 
and this training fails to optimize the electron density in some cases. 
This problem in the functional derivative leading to the erroneous solution of Euler equation is also reported in 
the previous efforts to develop KEDF with machine learning \cite{snyder-rupp-burke-prl12, snyder-rupp-burke-jcp13, yao-parkhill-jctc16, brockherde-naturecomm17, seino-nakai-jcp18,seino-nakai-cpl19,fujinami-nakai-cpl20}, where the training set is the KEDF itself. 
As the training and test sets in the deep learning, we adopt the {\it kinetic energy functional derivative} (KEFD) at each real-space grid point. 
It is noteworthy that the KEFD, [$\delta T [\rho] / \delta \rho] (\bm r)$, appears directly in the Euler equation Eq.~(\ref{Euler-ofdft}) and thus assures accuracy of the solution of the Euler equation, which is crucial to reproduce the target electron density $\rho (\bm r)$. 

The training set consists of the KEFD at 13,824 real-space grid points obtained for diamond. 
For efficient training, we adopt the stochastic natural gradient descent (SNGD), which is a mini-batch training based on natural gradient method\cite{amari-neural_comput98}. In SNGD, we randomly select $N_{b}$ training data, then update the $i$-th weight $W^{i}$ at $t$-th epoch as 
\begin{eqnarray}\label{WIter}
 W^{i}(t+1)&=&W^{i}(t) -\eta \sum_{k=1}^{n_{W}}[ G_{ik} 
 +\nu(t){\rm tr}({\bm G})\delta_{ik} ]^{-1} \nonumber \\
 & & \times\frac{\partial L}{\partial W^{k}}, 
\end{eqnarray}
where $\bm G$ is a metric tensor defined by 
\begin{equation}\label{def_metric}
 G_{ik}=\frac{1}{N_{b}}\sum_{p=1}^{N_{b}}
 \frac{\partial}{\partial W^{i}} \left[ \frac{\delta T^{\rm{NN}}({\bm r}_{p})}{\delta\rho}\right]
 \frac{\partial}{\partial W^{k}} \left[ \frac{\delta T^{\rm{NN}}({\bm r}_{p})}{\delta\rho}\right] . 
\end{equation}
Here the learning rate $\eta$ is fixed at $\eta=0.1$, and $n_{W}$ is the total number of the NN weight parameters, \{ $W^k$ \}. 
To achieve efficient convergence in Eq.~(\ref{WIter}), we blend the natural gradient and the ordinary gradient by introducing a weighted diagonal matrix $\nu(t){\rm tr}({\bm G})\delta_{jk}$, where the scheduling $\nu(t)=\nu_{0}/(1+bt)$ with $b=0.01$ is employed during the epochs being decreased from $\nu_{0}=10^{-5}$. The determination of those training parameters is explained in Appendix \ref{append:train_detail}.

The calculations of $\partial L / \partial {\bm W}$ and Eq.~(\ref{def_metric}) are performed by the backpropagation \cite{pukritt-neural_netw11,pukritt-jchemphys09}. The algorithm of the backpropagation in the present paper is briefly explained in  Appendix \ref{append:KEFD_train}. The training data set was divided into 90\% for the training and 10\% for the validation. 

\begin{table}[!h]
 \begin{center}
 \caption{Accuracy of our KEDF $\delta T^{\rm NN}/\delta \rho$ depending on the NN structures, i.e., 
 the number of hidden layers and the number of neurons per layer. The deviation from $\delta T^{\rm KS}/\delta \rho$ is represented in terms of the root mean square error (RMSE) in Ha (atomic unit of energy). 
}
 \begin{tabular}{cccc}\hline\hline
  \# of hidden & \# of neurons & \# of weight & \multirow{2}{*}{RMSE} \\
  layers & of per layer & parameters & \\ \hline
  1  & 5 & 21 &  0.356  \\ 
  1  & 10 & 41 & 0.310  \\
  1  & 15 & 61 & 0.302  \\
  1  & 20 & 81 & 0.306  \\
  2  & 5 & 51 & 0.304  \\
  2  & 10 & 151 & 0.311  \\
  3  & 5 & 81 & 0.296  \\
  \hline\hline
 \end{tabular}
 \label{tab_NN_struc}
 \end{center}
\end{table}

We have examined the dependence of the accuracy of the obtained $\delta T^{\rm NN}/\delta \rho$ on the structure of our NN, namely the number of hidden layers and the number of neurons in each layer. The accuracy is assessed by the deviation from the training set \{$\delta T^{\rm KS}/\delta \rho$\} at grid points in the real space. The 13824 grid points in a unit cell of diamond are chosen to construct the training set. Table~\ref{tab_NN_struc} shows the root mean square error (RMSE) of $\delta T^{\rm NN}/\delta \rho$ from $\delta T^{\rm KS}/\delta \rho$\ obtained in several NNs. 
Hereafter we abbreviate the atomic unit of energy (hartree) as Ha. 
From this examination, we have adopted the NN with three hidden layers and five neurons in each layer in the present paper. 
The actual weight parameters ${\bm W}$ of our NN is available upon request. 
The trend that the RMSE decrease with increasing the number of weight parameters observed in Table \ref{tab_NN_struc} is indicative of further improvement of our NN for better KEDF in future. 

\subsection{Augmentation of the enhancement factor}\label{ssec:augmentation}

The enhancement factor in the present paper is further augmented by requiring rigorous limits for $s \to 0$ and $s \to \infty$: When $s \to 0$, it should be the second-order gradient expansion \cite{brack-jennings-chu-physlett76} as $1+(5/27)s^{2}$, whereas when $s \to \infty$ it should be equal to the vW KEDF \cite{vweizsacker-zphys35}, $F \to (5/3)s^{2}$. A form which satisfies those limits within the first order of $q^2$ is, 
\begin{equation}\label{def_PGSL}
 F^{(0)}(s^{2},q)=\frac{5}{3}s^{2}+e^{-\alpha s^{2}}+\beta q^{2}, 
\end{equation}
with $\alpha$ being 40/27. 

Here, $\beta$ was previously adjusted to reproduce structural properties of target materials \cite{constantin-PGSL-jpclett18,constantin-PGSL-jctc19}. 
Instead, we here determine $\beta$ so that the second functional derivative of KEDF derived from Eq.~(\ref{def_PGSL}) in the small-$s$ limit reproduces the Lindhard function\cite{lindhard}. We have found that $\beta=0.382$ well satisfies this homogeneous-limit condition (for the fitting procedure, see Appendix~\ref{append:F0_fit}).

Then we finally propose our enhancement factor, 
\begin{equation}\label{def_NN_sub}
 \tilde{F}^{\rm{NN}}(s^{2},q)=X(q)F^{(0)}(s^{2},q)+[1-X(q)]F^{\rm{NN}}(s^{2},q), 
\end{equation}
where $X(q) = \exp(-A q^{4})$ is an interpolation function between the small-$q$ and large-$q$ subsystems. 
This is to enhance the accuracy especially for large $q$ obtained by the flexibility of the NN, following the concept of subsystem functionals in DFT \cite{kohn-mattson-prl98,mattson-kohn-jchemphys01,armiento-mattson-prb02,mattson-armiento-intjqchem10}. 
Our final KEDF, $\tilde{T}^{\rm{NN}}$, 
is given by Eq.~(\ref{NN_enhance_fac}) with ${F}^{\rm{NN}}$ being replaced by $\tilde{F}^{\rm{NN}}$. 

The parameter $A$ can be optimized by minimizing the cost function $L$ by expanding the metric tensor $G_{jk}$ to $(n_W + 1) \times (n_W + 1) $ dimensions by adding the components related to $\partial [{\delta \tilde{T}^{\rm{NN}}({\bm r}_{p})}/{\delta\rho}]/\partial A$. The optimized $A$, however, shows multi-minima structure with nearly flat dependence of $L$ on $A$ in the range $10<A<32$, causing uncertainty in the optimization. Then we additionally minimize the difference between the electron density from our KEDF and the KS electron density, which more reliably settles $A$ at 31.62 (see Appendix \ref{append:A_opt} for details).

\section{Results and Discussion}\label{sec:results}

\subsection{Behavior of NN kinetic energy density functional}\label{ssec:kedf_plot} 

\begin{figure}
\centering
\includegraphics[width=.85\linewidth]{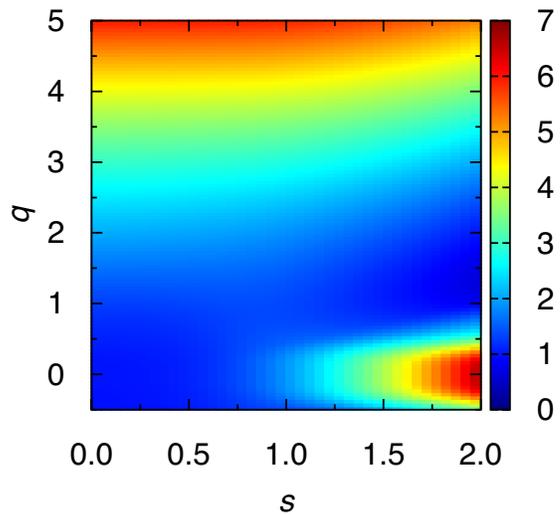}
\caption{Contour color map of the enhancement factor NN KEDF $\tilde{F}^{\rm{NN}}(s^2,q)$ plotted as a function of $s$ and $q$ 
within a typical range of $s$ and $q$ ($0\le s \le 2.0 \;, -0.5\le q \le 5.0$). 
The abscissa is the reduced density gradient $s$ and the ordinate is the reduced Laplacian of density $q$.
}\label{NNsub_2d_map}
\end{figure}

\begin{figure}
\centering
\includegraphics[width=.9\linewidth]{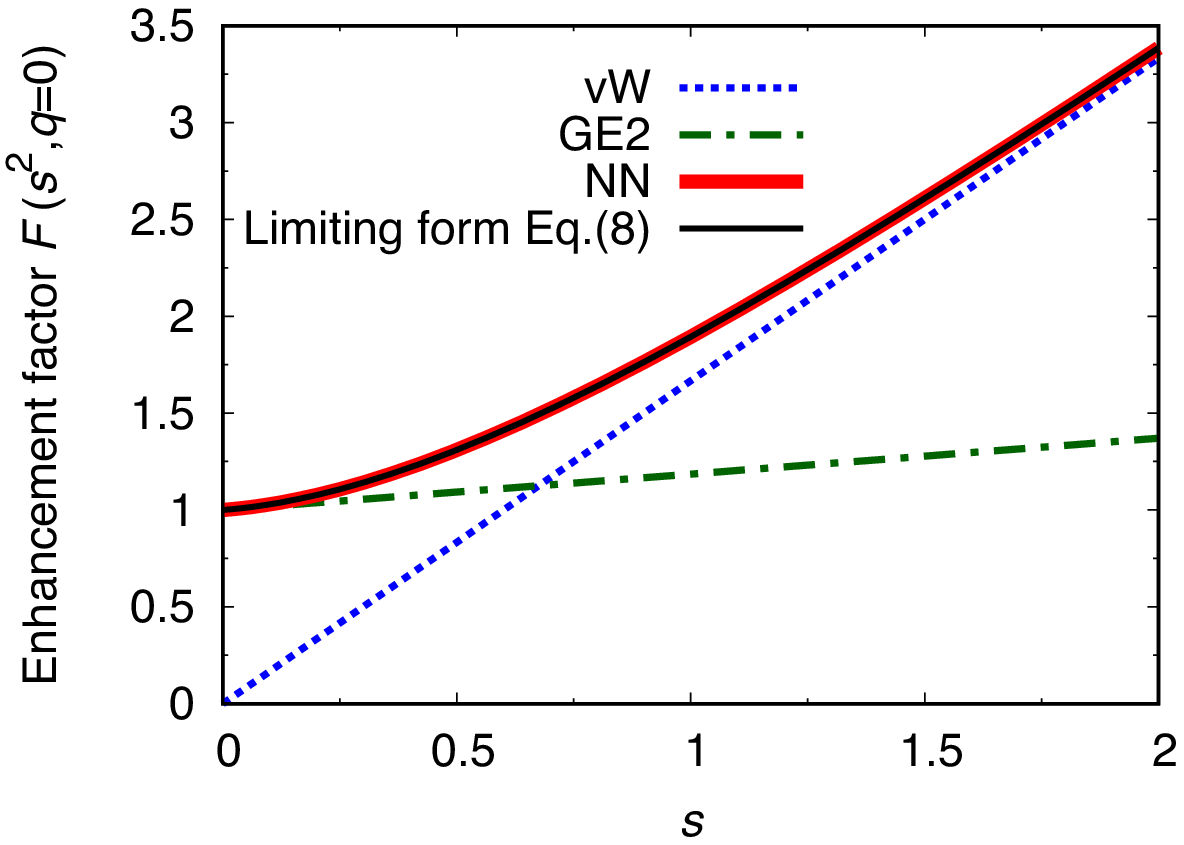}
\caption{$s$-dependence of the enhancement factors at $q=0$ 
within a typical range of $s$ ($0\le s \le 2.0$). 
The abscissa is the reduced density gradient $s$ and the enhancement factor $F(s^2,q=0)$. 
vW, GE2, and NN denote the enhancement factors of the vW KEDF\cite{vweizsacker-zphys35}, the second-order gradient expansion\cite{brack-jennings-chu-physlett76}, and $\tilde{F}^{\rm{NN}}$ (the present paper Eq.~(\ref{def_NN_sub})), respectively.
}\label{F1d_q0}
\end{figure}

\begin{figure}
\centering
\includegraphics[width=.9\linewidth]{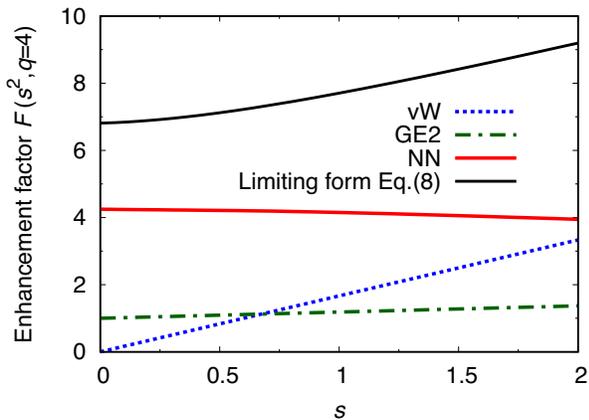}
\caption{$s$-dependence of the enhancement factors at $q=4$ 
within a typical range of $s$ ($0\le s \le 2.0$). 
The abscissa is the reduced density gradient $s$ and the enhancement factor $F(s^2,q=4)$. 
vW, GE2, and NN denote the enhancement factors of the vW KEDF\cite{vweizsacker-zphys35}, the second-order gradient expansion\cite{brack-jennings-chu-physlett76}, and  $\tilde{F}^{\rm{NN}}$ (the present paper Eq.~(\ref{def_NN_sub})), respectively.
}\label{F1d_q4}
\end{figure}

Figure \ref{NNsub_2d_map} shows our NN enhancement factor $\tilde{F}^{\rm{NN}}(s^2,q)$, Eq.~(\ref{def_NN_sub}), for a typical range of $s$ and $q$. 
It is clear that the NN enhancement factor is continuous in $(s^2,q)$-plane, indicating that $\tilde{F}^{\rm{NN}}$ smoothly connects ${F}^{\rm{NN}}$, Eq.~(\ref{NN_KEDF_explicit}), and the limiting form $F^{(0)}$, Eq.~(\ref{def_PGSL}). To compare $\tilde{F}^{\rm{NN}}$, $F^{(0)}$, $F^{\rm vW}=(5/3)s^{2}$ (the vW KEDF \cite{vweizsacker-zphys35}), and also the second-order gradient expansion $F^{\rm GE2}=1+(5/27)s^{2}$ \cite{brack-jennings-chu-physlett76}), we plotted the $s$-dependence of these enhancement factors for two fixed values of $q$, namely $q=0$ and $q=4$ as the typical large $q$-value in most systems, in Figs.~\ref{F1d_q0} and \ref{F1d_q4}. At $q=0$, $\tilde{F}^{\rm{NN}}$ recovers the limiting form $F^{(0)}$ as expected. The behavior of $\tilde{F}^{\rm{NN}}$ is substantially different from the limiting form $F^{(0)}$ at $q=4$, which is ascribed to the form of NN KEDF, Eq.~(\ref{NN_KEDF_explicit}), and relevant to its accuracy.

\subsection{Accuracy of NN KEDF}\label{ssec:accuracy}

\subsubsection{Electron densities}\label{sssec:rho}
We assess the accuracy of KEDFs with the self-consistent-field (SCF) density obtained by minimizing the total energy with the developed KEDF. 
One way is the direct minimization with iterative techniques\cite{ho-ligneres-carter-cpc08,hung-huang-cater-cpc10,chen-xia-cater-cpc15,mi-shao-su-cpc16}. The other way is to solve the Euler equation Eq.~(\ref{Euler-ofdft}) by the matrix diagonalization, which we adopt in this work. By introducing vW KEDF \cite{vweizsacker-zphys35}, $T_{\rm vW} [\rho]$ which satisfies $\delta T_{\rm vW} / \delta \rho = - (\nabla^2 \sqrt{\rho}) / (2 \sqrt{\rho})$, the Euler equation becomes a Schr\"odinger-type equation for $ \sqrt{\rho} $ \cite{levy-perdew-sahni-pra84}. However, it is recognized that the diagonalization of such Schr\"odinger-type equation 
suffers from ill convergence~\cite{karasiev-trickey-cpc12}. 
We have found that this difficulty can be circumvented in all KEDFs we considered 
by introducing a parameter $\lambda$ and by performing the rearranged diagonalization scheme. 
Previously, this sort of rearrangement has been applied to the TF$(\lambda )$vW functional 
with $\lambda = 1/5$ or 1/9 \cite{lehtomaki-makkonen-caro-jcp14}. 
We first express KEDF as the sum of scaled vW functional and the rest part: 
\begin{equation}
 T_{s}[\rho] = \lambda T_{\rm{vW}}[\rho] + T_{r}[\rho] .
\end{equation}
Then Eq.~(\ref{Euler-ofdft}) becomes 
\begin{equation}\label{scheq-ofdft}
   \left[ -\frac{\nabla^2}{2} + \frac{1}{\lambda} 
   \left( 
   \frac{\delta T_r [\rho]}{\delta \rho} (\bm r)  
   + v_{\rm{KS}}([\rho];{\bm r}) 
   \right) 
   \right]  \sqrt{\rho({\bm r})}=\frac{\mu}{\lambda}\sqrt{\rho({\bm r})},
\end{equation} 
where $v_{\rm{KS}} = v_{\rm ext} + \delta E_{\rm H} / \delta \rho + \delta E_{\rm xc} / \delta \rho$ is the usual KS potential. 
In our scheme, we have found that $\lambda=10$ leads to 
a satisfactory convergence in the SCF calculations. 
Note that any choice of $\lambda$ offers the same correct SCF solution, if it converges.

\begin{figure}
\centering
\includegraphics[width=.9\linewidth]{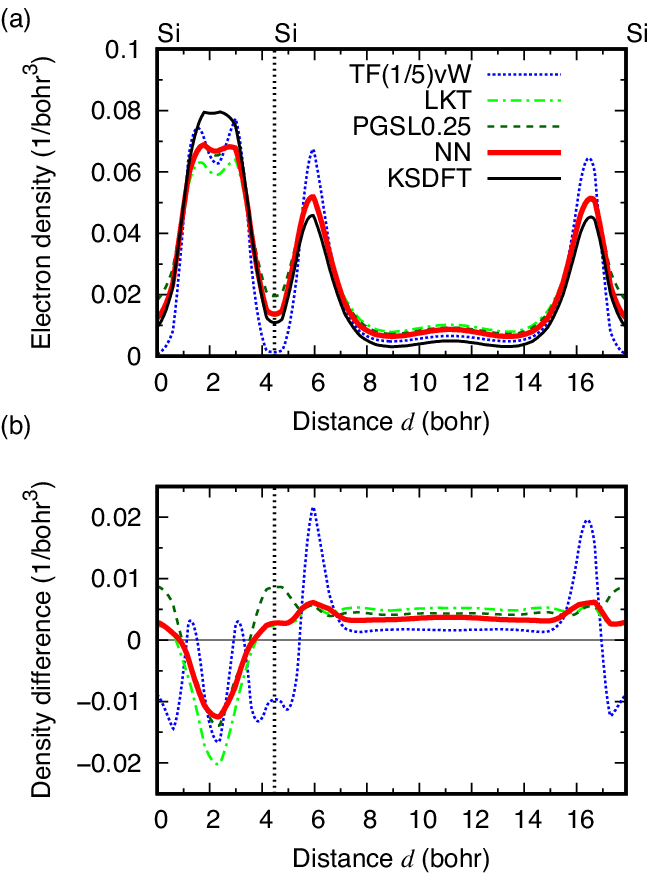}
\caption{(a) SCF electron density in ds-Si along $[111]$ direction obtained by different approximations to KEDF. 
The horizontal axis is the distance $d$ from a Si atom along $[111]$. 
Si atoms are located at the positions depicted by the vertical dashed line, and the left and right ends. 
(b)  Difference of densities with respect to the KS density (black solid line in (a)). The 0 value on the ordinate is indicated by a horizontal black solid line.
}\label{rho_compare}
\end{figure}

\begin{table*}
 \begin{center}
 \caption{RMSE of SCF density in $10^{-2} \times {\rm bohr}^{-3}$ with respect to the KS density in the periodic systems 
obtained by 5 different KEDFs. The right end column (``ratio") indicates the ratio of RMSE averaged over all the systems to the averaged RMSE obtained by NN KEDF.}
 \begin{tabular}{l |ccccccccccc|c} \hline\hline
      & diamond & graphene & ds-Si & fcc-Si & $\beta$-tin Si & 3C-SiC & bcc-Li & fcc-Al & fcc-Cu & bcc-Na & NaCl & ratio \\ \hline
NN & 1.1450 & 0.6467 & 0.3580 & 0.1242 & 0.2370 & 0.6180 & 4.1915 & 0.0971 & 8.4183 & 2.8791 & 2.1998 & 1\\
NN$^{[\rm{bare]}}$ & 0.9370 & 0.4807 & 0.3050 & 0.0889 & 0.2120 & 0.6110 & 2.1484 & 0.0800 & 8.7775 & 1.6449 & 1.2761 & 0.777\\
PGSL0.25 & 1.2850 & 0.7097 & 0.4350 & 0.1583 & 0.2580 & 0.7140 & 5.3014 & 0.1205 & 11.193 & 4.1229 & 3.4720 & 1.255 \\
LKT & 1.6760 & 0.6665 & 0.5110 & 0.1113 & 0.3120 & 0.9630 & 5.5690 & 0.1130 & 12.866 & 4.3014 & 3.7826 & 1.357 \\
TF(1/5)vW & 2.0560 & 0.3151 & 0.5500 & 0.5755 & 0.4850 & 1.2010 & 5.7562 & 0.3488 & 25.770 & 5.0822 & 3.1791 & 2.153 \\ \hline\hline
 \end{tabular}
 \label{tab_rmse_rho_bulk}
 \end{center}
\end{table*}

\begin{table*}
 \begin{center}
 \caption{RMSE of SCF density in $10^{-2} \times {\rm bohr}^{-3}$ with respect to the KS density in atoms. The right column (``ratio") indicates the ratio of RMSE to NN averaged over all systems.}
 \begin{tabular}{l |ccccccc|c} \hline\hline
      & Li & C & Na & Al & Si & Cl & Cu & ratio \\ \hline
NN & 1.1155 & 0.1469 & 1.2508 & 0.0271 & 0.0481 & 0.2943  & 2.2284 & 1\\
NN$^{[\rm{bare]}}$ & 0.7003 & 0.1253 & 0.9795 & 0.2322 & 0.0415 & 0.2250 & 2.7214 & 1.9544  \\
PGSL0.25 & 1.6069 & 0.3482 & 1.7854 & 0.2686 & 0.1527 & 0.4664 & 4.4330 & 3.2613 \\
LKT & 2.1197 & 0.2540 & 1.6340 & 0.0532 & 0.0776 & 0.4145 & 5.0251 & 1.7393 \\
TF(1/5)vW & 1.3733 & 0.1315 & 1.2050 & 0.0329 & 0.0635 & 0.2832 & 1.3952 & 1.0302 \\ \hline\hline
 \end{tabular}
 \label{tab_rmse_rho_atom}
 \end{center}
\end{table*}

\begin{table}
 \begin{center}
 \caption{RMSE of SCF density in $10^{-2} \times {\rm bohr}^{-3}$ with respect to the KS density in diatomic molecules. The right column (``ratio") indicates the ratio of RMSE to NN averaged over all systems.}
 \begin{tabular}{l |cccccc|c} \hline\hline
      & Li$_{2}$ & C$_{2}$ & Na$_{2}$ & Al$_{2}$ & Si$_{2}$ & Cl$_{2}$ & ratio \\ \hline
NN & 0.7036 & 0.3545 & 1.4726 & 0.2066 & 0.1000 & 0.5534  & 1\\
NN$^{[\rm{bare]}}$ & 0.4036 & 0.2258 & 1.3239 & 0.2137 & 0.1047 & 0.4204 &  0.8250  \\
PGSL0.25 & 0.6932 & 0.3919 & 1.6260 & 0.1882 & 0.2661 & 0.5130 & 1.2823 \\
LKT & 0.7616 & 0.6422 & 2.4138 & 0.1548 & 0.2108 & 0.7267 & 1.4505 \\
TF($\frac{1}{5}$)vW & 0.4342 & 0.7200 & 3.8073 & 0.1618 & 0.2019 & 0.4080 & 1.4620 \\ \hline\hline
 \end{tabular}
 \label{tab_rmse_rho_dimer}
 \end{center}
\end{table}

Figure~\ref{rho_compare} shows calculated SCF electron density obtained by our NN KEDF. 
For comparison, the computed densities using the KEDFs in the past, i.e., PGSL0.25\cite{constantin-PGSL-jpclett18} defined by Eq.~(\ref{def_PGSL}) with $\beta=0.25$, LKT\cite{luo-LKT-prb18}, and the conventional TF(1/5)vW (TF$(\lambda)$vW KEDF with $\lambda=1/5$) \cite{berk-pra83} are also shown. Figure~\ref{rho_compare} (a) is the electron density along $[111]$ direction in diamond-structured(ds)-Si 
and Fig.~\ref{rho_compare} (b) shows its difference from the KS density. 
The obtained electron density shows overall superiority of 
our NN KEDFs against PGSL0.25 and TF(1/5)vW most clearly visible at the nuclear site ($d=4.47$ bohr). When compared with LKT, NN tends to be more accurate in the bonding region ($0<d<4.47$ bohr). 
We have also calculated the electron densities with our NN KEDF of other 23 systems (10 solids, 6 diatomic molecules and 7 atoms), and found that the obtained densities satisfactorily reproduce the corresponding KS densities (Supplemental Material \cite{suppl_mater} for details). It is noteworthy that our NN KEDF obtained by the training data of only ds-C shows high transferability for various systems.

The superiority of our NN KEDF in reproducing the KS densities is quantified by the RMSE of the SCF density $\rho_{\rm scf}$ with respect to the KS density $\rho^{\rm KS}$ for all the 24 systems examined in the present paper. Table~\ref{tab_rmse_rho_bulk} shows the RMSE for periodic systems, including semiconductors, metals and an ionic material with some of their polytypes: 
diamond, graphene, ds-Si, face-centered-cubic(fcc)-Si, $\beta$-tin Si, zincblende(3C)-SiC, body-centered-cubic(bcc)-Li, fcc-Al, fcc-Cu, bcc-Na, and NaCl. 
For comparison, the results obtained with the KEDFs in the past are also shown. We also present the results by the uninterpolated NN enhancement factor $F^{\rm{NN}}$ (i.e., $A \to \infty$ in Eq.~(\ref{def_NN_sub})) (labelled as NN$^{[\rm{bare}]}$) trained by the same scheme as NN.
The overall good performance of NN is demonstrated by the RMSE for each material. The RMSE averaged over all 11 materials (the right end column ``ratio" in Table~\ref{tab_rmse_rho_bulk}) clearly shows the superiority of the present NN KEDF to other KEDFs in the past. 

The RMSE of the SCF density has been also computed for 13 isolated systems, 
Li, C, Na, Al, Si, Cl, and Cu atoms, 
and Li$_{2}$, C$_{2}$, Na$_{2}$, Al$_{2}$, Si$_{2}$, and Cl$_{2}$ molecules, 
and then averaged over all the atoms and molecules, as shown in Table \ref{tab_rmse_rho_atom} and \ref{tab_rmse_rho_dimer}.
The same RMSE ``ratio" indicates that NN and TF(1/5)vW are most accurate in atoms, whereas NN and NN$^{[\rm{bare}]}$ are most accurate in molecules.

These results indicate that we have  succeeded to construct NN KEDFs that outperform previous ones without resorting to system-dependent parameters. 
The electron density calculated with NN$^{\rm [bare]}$ shows better performance than NN in some cases. However, physical quantities shown below indicate the limitation of NN$^{\rm [bare]}$. The present NN up to $\ell_1=1$ makes the agreement with KS at the first-order derivative of $T$, while the density is a quantity related at the first order level $\rho({\bm r})=-\delta E[\rho]/\delta \mu({\bm r})$ where $\mu({\bm r})$ is the local chemical potential. Hence the success in reproducing the KS density is intrinsic to the present scheme.

The NN adopted in the present paper produces the minimum value of the cost function $L_{\rm min} = 3.519 \times 10^{-2}$ Ha$^2$. We have examined other NNs which produce the cost function of $L_{\rm min} < L \le 1.8L_{\rm min}$ Ha$^2$. We have confirmed that the obtained RMSE of the electron density with such NN increases less than 1\% at most. 

\subsubsection{Structural properties}\label{sssec:struc}

To further examine the validity of our OFDFT scheme with the NN KEDF, we have calculated structural properties and energetics for test systems: the lattice constants $a_0$, bulk moduli $B_0$ and cohesive energies $E_{\rm coh}$ of 11 solids, and the bond lengths $r_0$ of 6 molecules. 

Tables~\ref{tab_a0_B0_kedf1} and \ref{tab_a0_B0_kedf2} show the lattice constants and bulk moduli for 11 solids along with the corresponding values obtained using other KEDFs in the past. 
We compare the relative errors with respect to the KS values (numbers in the parentheses in Tables~\ref{tab_a0_B0_kedf1} and \ref{tab_a0_B0_kedf2}). 
Our NN KEDF provides the smallest relative errors in 10 cases ($a_0$ for $\beta$-tin Si, fcc-Al, fcc-Cu, graphene; $B_0$ for diamond, fcc-Si, $\beta$-tin Si, fcc-Al, fcc-Cu, NaCl), whereas the number of the cases with the smallest errors are 6 for LKT and 5 for PGSL0.25. 
The overall quantitative index of the superiority is evaluated as the mean absolute relative errors (MAREs) with respect to the KS values for those quantities. 
The NN KEDF clearly shows the minimum MAREs for both $a_0$ and $B_0$ (Table~\ref{tab_a0_B0_kedf2}), indicating its superior performance. 
For all structural properties, NN$^{[\rm bare]}$ produces larger MAREs than the NN functional. 
This indicates the importance of augmenting the enhancement factor and the validity of concept of the subsystem DFT \cite{kohn-mattson-prl98}. 
It is noteworthy that our NN KEDF functional is trained by a part of the data in diamond and reproduces KSDFT results reasonably for the structural properties of a variety of materials including metals and ionic crystals 
with no adjustable parameters, where the machine learning parameters are determined uniquely in an {\it ab initio} fashion after minimizing the cost function.

Table~\ref{tab_r0_kedf} shows the bond lengths of 6 molecules. 
The MAREs with respect to the KS values show that NN KEDF performs overall better than other KEDFs. 
However, the obtained MAREs of the bond lengths of the molecules are certainly larger than the MAREs of the lattice constants of the solids. This is presumably due to the fact that we have trained our NN using the data of the solid, diamond-structured carbon. This demonstrates the difficulty to develop the KEDF quantitatively valid for both localized and delocalized systems. 
Table~\ref{tab_Ecoh_kedf} shows the calculated cohesive energies ($E_{\rm coh}$) for 11 solids. The chemical trend obtained in the KS scheme is reproduced by our NN KEDF. However, the quantitative reproduction of the KS values is not satisfactory. In fact, all the KEDFs including those in the past provide MAREs of the cohesive energy larger than 20 \% (the right-end column). 
Among them, the NN KEDF keeps the smallest MAREs. 

The lattice constant and the bulk modulus are obtained by the behavior of the total energy $E$ as a function of the volume $V$ around its minimum point: $a_{0}$ is obtained as the zero point of $\delta E/\delta V$ and $B_{0}$ requires the second derivative $\delta^2 E/\delta V^2$. In our OFDFT with the NN KEDF with $\ell_1=1$, the first derivative of $E$ with respect to the density is trained. 
Hence the high reproducibility of $a_0$ may be intrinsic in the present scheme, while it is natural that the MAREs for $B_0$ are worse than those for $a_0$. 
Along this line, the machine learning up to $\ell_1=2$ is desired for physical quantities associated with the second derivative of $E$ such as $B_0$. 
Since $B_0$ calculated with $\ell_1=1$ already shows fair agreements with that by KS-DFT and experiments, the higher-order machine learning has potential to provide us with systematically more accurate KEDF.

\begin{table*}
 \begin{center}
 \caption{Comparison of equilibrium lattice constant ($a_{0}$) in {\AA} and bulk moduli ($B_{0}$) in GPa obtained by 
NN and NN$^{\rm [bare]}$ with those obtained by different approximations to KEDF. 
Numbers in the parentheses are the relative errors in \% with respect to the KS values. 
 In $\beta$-tin Si, the ratio of another lattice constant $c_{0}$ to $a_{0}$ ($c_{0}/a_{0}$) is also listed. 
 Some values are left as blanks because the total energy monotonically decreases with respect to the volume expansion.}
 \begin{tabular}{l |cccccccccc ccccc} \hline\hline
                                      & \multicolumn{2}{c}{diamond} & \multicolumn{2}{c}{ds-Si} & \multicolumn{2}{c}{fcc-Si} & \multicolumn{3}{c}{$\beta$-tin Si} & \multicolumn{2}{c}{3C-SiC} & \multicolumn{2}{c}{bcc-Li} & \multicolumn{2}{c}{fcc-Al} \\ 
                                      & $a_{0}$ & $B_{0}$ & $a_{0}$ & $B_{0}$ & $a_{0}$ & $B_{0}$  & $a_{0}$ & $c_{0}/a_{0}$ & $B_{0}$ &  $a_{0}$ & $B_{0}$ & $a_{0}$ & $B_{0}$ & $a_{0}$ & $B_{0}$  \\ \hline
 \multirow{2}{*}{NN}                             & 3.428  & 411    & 5.367   & 85.4    & 3.642  & 143  & 4.673 & 0.529  & 165    & 3.092 & 157    &3.479   & 15.5   & 4.118 & 79.1  \\
                                                          & (-2.53)  & (25.7) & (-0.46) & (-14.6) & (-0.22) & (5.93) & (0.17) & (-1.67) & (5.10) & (2.28)  & (-36.2) & (-0.83) & (4.03) & (1.68) & (2.06) \\ 
 \multirow{2}{*}{NN$^{[\rm{bare]}}$}   & 3.130    & 867   & 5.187 & 83.5    & 3.529    & 149    & --     & --        & --         & 2.904    & 253  & 3.329   &  18.9 & 3.841 & 129   \\
                                                         & (-11.0) & (165) &(-3.80) & (-16.5) & (-3.32) & (10.4)      &      &            &          & (-3.94) & (2.85) & (-5.10) & (26.8)   & (-5.16) & (66.5) \\
 \multirow{2}{*}{PGSL0.25}                 & 3.430  & 433    & 5.384  & 93.4 & 3.702    & 118       & 4.744 & 0.529        & 137   & 3.073   & 217 & 3.496  &  15.5      & 4.197 & 67.4 \\
                                                         & (-2.47) & (32.4) & (-0.15) & (-6.6)  & (1.42) & (-13.0)    & (1.69) &(-1.67)        & (-12.7) & (1.65) & (-11.8) & (-0.33) & (3.78) & (3.62) & (-13.0)  \\
 \multirow{2}{*}{LKT}                          & 3.343 & 578    & 5.336  & 104    & 3.644 &  161   & 4.643 &0.532 &  186     & 3.066 &  227    &  3.492 & 15.3   & 4.144 & 86.0\\
                                                         & (-4.95) & (76.8) & (-1.04) & (4.00) & (-0.16) & (19.5) & (-0.47) &   (-1.12)  & (18.5) & (1.42) & (-7.72) & (-0.46) & (2.52) & (2.32) & (10.9)  \\
 \multirow{2}{*}{TF(1/5)vW}                & -- & -- & 5.708  & 39.6    & 3.847 & 54.0   & --    & -- & --                                 & 3.353    & 63.2  & 3.384 & 16.4    & 4.243 & 44.0 \\
                                                         &    &    & (5.86)  & (-60.4) &(5.40) & (-60.0) &       &     &                                      &   (10.9) & (-74.3)     & (-3.54) & (10.1) & (4.76) & (-43.2) \\ \hline
 KSDFT                                              & 3.517 & 327 & 5.392 & 100 & 3.650 & 135 & 4.665 & 0.538 & 157 & 3.023 & 246 & 3.508 & 14.9 & 4.050 & 77.5   \\ 
\hline\hline
 \end{tabular}
 \label{tab_a0_B0_kedf1}
 \end{center}
\end{table*}

\begin{table*}
\begin{center}
 \caption{Comparison of equilibrium lattice constant ($a_{0}$) in {\AA} and bulk moduli ($B_{0}$) in GPa  obtained by 
NN and NN$^{\rm [bare]}$
with different approximations to KEDF. 
 Numbers in the parentheses are the relative errors in \% with respect to the KS values. 
 Mean absolute relative errors (\%) with respect to the KS values averaged over the systems including the ones in Table~\ref{tab_a0_B0_kedf1} are also listed as MARE.}
\begin{tabular}{l|ccccccc|cc} \hline\hline
                                      & \multicolumn{2}{c}{fcc-Cu} & \multicolumn{2}{c}{bcc-Na} & \multicolumn{2}{c}{NaCl} & graphene & \multicolumn{2}{c}{MARE} \\ 
                                      & $a_{0}$ & $B_{0}$ & $a_{0}$ & $B_{0}$ & $a_{0}$ & $B_{0}$  & $a_{0}$ & $a_{0}$ & $B_{0}$ \\ \hline
 \multirow{2}{*}{NN}                            &  3.730 & 169 & 4.227 & 7.98 & 5.678 & 27.7 & 2.448 &  &  \\
                                                         &  (2.33) & (6.96) & (-0.91) & (3.50) & (3.61) & (6.94) & (0.29) & 1.39 & 11.1   \\
 \multirow{2}{*}{NN$^{[\rm{bare]}}$}  & 3.887 & 228  & 4.045 & 9.73  & 5.187  & 32.4  & 2.377   &  &  \\
                                                        & (6.64) & (44.3) &(-5.18) & (26.2) & (-5.35) & (25.1) & (-2.62) & 4.74 & 38.4 \\
 \multirow{2}{*}{PGSL0.25}                & 3.795 & 138   & 4.250 & 8.00   & 5.595  & 22.9   & 2.433   &  &  \\
                                                        & (4.12) & (-12.4) &(-0.38) & (3.75) & (2.10) & (-11.5) & (-0.32)  & 1.66 & 12.1  \\
 \multirow{2}{*}{LKT}                        & 3.762 & 175   & 4.245 & 7.91   & 5.596 & 23.8  & 2.402   &  &   \\
                                                       & (3.21) & (10.7) &(-0.50) & (2.61) & (2.12) & (-8.20) & (-1.59)  & 1.66 & 14.5 \\
 \multirow{2}{*}{TF(1/5)vW}              & 3.799 & 88.4  & 4.116 & 8.49   & 6.056  & 6.75   & 2.593 &  &    \\
                                                       & (4.21) & (-44.1) &(-3.51) & (10.1) & (10.5) & (-73.9) & (6.24) & 6.10 & 47.0   \\ \hline
 KSDFT                                            & 3.645 & 158   & 4.266 & 7.71   & 5.480  & 25.9   & 2.441   & &  \\
 \hline\hline
 \end{tabular}
 \label{tab_a0_B0_kedf2}
 \end{center}
\end{table*}

\begin{table*}
 \begin{center}
 \caption{Comparison of the equilibrium bond length ($r_{0}$) in {\AA} for molecules. Numbers in the bottom parentheses are the relative errors in \% with respect to the KS values. 
 Mean absolute relative errors in \% with respect to the KS values are also listed as MARE.}
 \begin{tabular}{l |cccccc|c} \hline\hline
      & Li$_{2}$ & C$_{2}$ & Na$_{2}$ & Al$_{2}$ &Si$_{2}$ & Cl$_{2}$ & MARE \\ \hline
\multirow{2}{*}{NN} & 3.118  & 1.425 & 2.913 & 2.608 & 2.325 & 1.966  \\
                              & (8.04) & (18.5) & (-5.56) & (-4.32) & (5.26) & (5.20)  & 7.81 \\
\multirow{2}{*}{NN$^{[\rm{bare]}}$} & 2.332 & 1.414 & 2.995 & 2.886 & 2.243 & 1.964   \\
                              & (-19.2) & (17.6) & (-2.90) & (5.90) & (1.54) & (3.54) & 8.45 \\
\multirow{2}{*}{PGSL0.25} & 2.970  & 1.388 & 3.369 & 2.815 & 2.497 & 1.966  \\
                              & (2.91) & (15.4) & (9.23) & (3.31) & (13.0) & (3.62) & 7.92 \\
\multirow{2}{*}{LKT} & 2.846  & 1.391 & 3.211 & 3.153 & 1.974 & 1.777   \\
                              & (-1.37) & (15.7) & (4.12) & (15.7) & (-10.6) & (-6.32) & 8.96 \\
 \multirow{2}{*}{TF(1/5)vW} & 3.011  & 1.489 & 3.042 & 2.505 & 2.582 & 2.020      \\
                              & (4.33) & (23.9) & (-1.37) & (-8.09) & (16.9) & (6.49) & 10.2 \\ \hline
KSDFT & 2.886  & 1.203 & 3.084 & 2.725 & 2.209 & 1.897 & \\
\hline\hline
 \end{tabular}
 \label{tab_r0_kedf}
 \end{center}
\end{table*}

\begin{table*}
 \begin{center}
 \caption{Comparison of cohesive energies ($E_{\rm coh}$) for solids in Ha. Numbers in the parentheses are the relative errors in \% with respect to the KS values. The mean absolute relative errors (MAREs) in \% with respect to the KS values are also listed.}
 \begin{tabular}{l |ccccccccccc|c} \hline\hline
      & diamond & graphene & ds-Si & fcc-Si & $\beta$-tin Si & 3C-SiC & bcc-Li & fcc-Al & fcc-Cu & bcc-Na & NaCl & MARE  \\ \hline
\multirow{2}{*}{NN} & -0.3544  & -0.3038 & -0.3714 & -0.2036 & -0.3710 & -0.3445 & -0.0668 & -0.1258 & -0.1073 & -0.0443 & -0.3068 &   \\
                              & (22.1) & (40.2) & (14.8) & (8.55) & (15.2) & (32.2) & (7.1) & (19.1) & (16.6) & (21.0) & (22.9) & 20.0  \\
\multirow{2}{*}{NN$^{[\rm{bare]}}$} & -0.3517  & -0.2332 & -0.3731 & -0.2416 & -0.4186 & -0.3341 & -0.0926 & -0.5550 & -0.1786 & -0.0692 & -0.4944 &  \\
                              & (22.7) & (54.1) & (14.4) & (8.50) & (4.38) & (34.3) & (48.4) & (425) & (94.0) & (89.0) & (98.1) & 81.2 \\
\multirow{2}{*}{PGSL0.25} & -1.2546  & -0.7595 & -0.7839 & -0.4581 & -0.8854 & -1.0818 & -0.2079 & -0.6102 & -0.2471 & -0.0992 & -0.6907 &  \\
                              & (176) & (49.6) & (79.8) & (106) & (102) & (113) & (233) & (477) & (168) & (171) & (177) & 168 \\
\multirow{2}{*}{LKT} & -1.1272  & -0.5836 & -0.6768 & -0.4422 & -0.8340 & -0.9147 & -0.2134 & -0.2770 & -0.1856 & -0.0771 & -0.5235  &    \\
                              & (148) & (14.9) & (55.2) & (98.6) & (90.5) & (79.9) & (242) & (162) & (102) & (111) & (110) & 110 \\
 \multirow{2}{*}{TF(1/5)vW} & -0.3860  & -0.1877 & -0.3634 & -0.2187 & -0.4148 & -0.4257 & -0.0664 & -0.1369 & -0.1158 & -0.0499 & -0.3484 &     \\
                              & (15.2) & (137) & (16.7) & (1.77) & (5.25) & (16.3) & (6.51) & (29.5) & (25.8) & (36.3) & (39.6) & 30.0  \\ \hline
KSDFT & -0.4551  & -0.5077 & -0.4361 & -0.2226 & -0.4377 & -0.5084 & -0.0624 & -0.1057 & -0.0921 & -0.0366 & -0.2496 \\\hline\hline
 \end{tabular}
 \label{tab_Ecoh_kedf}
 \end{center}
\end{table*}

\subsection{Order $N$ DFT computations}\label{ssec:scaling}

Finally, we have analyzed the computational time of both KSDFT and OFDFT in order to demonstrate our $\mathcal{O}(N)$ scheme. 
We decompose the computational time for a single SCF iteration ($t_{\rm{SCF}}$) as $t_{\rm{SCF}}=t_{1}+t_{2}+t_{3}+t_{\rm{others}}$, where $t_{1}$ is the time for the subspace diagonalization (SD), conjugate-gradient minimization (CG) 
and Gram-Schmidt (GS) orthonormalization, $t_{2}$ for updating density and calculating the total energy, $t_{3}$ for the mixing procedure to obtain the new input potential, and $t_{\rm{others}}$ for other procedures such as MPI gathering and broadcasting eigenvectors. The test systems are 4H-SiC supercells with sizes of 576, 1024, 1600, 2400, and 4704 atoms. Only gamma point is sampled for the Brillouin zone integration in KSDFT. The grid-spacing is chosen as 0.39 {\AA}. We use 36 eigenvectors for the diagonalization of 
Eq.~(\ref{scheq-ofdft}). 

\begin{figure}
\centering
\includegraphics[width=.95\linewidth]{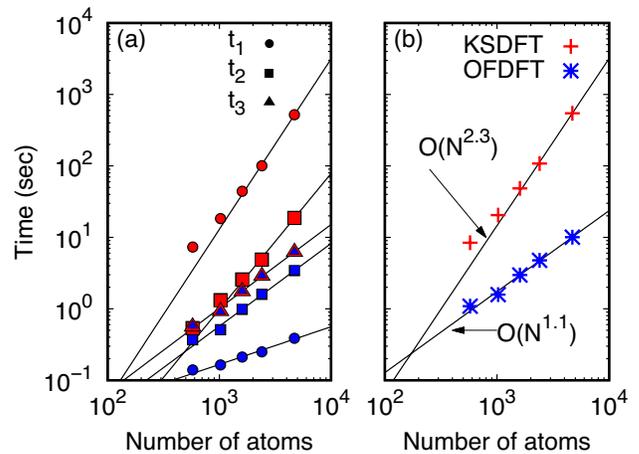}
\caption{Computational time of KSDFT and OFDFT. 
Comparison of (a) $t_{1}$, $t_{2}$, and $t_{3}$; 
and (b) $t_{\rm{SCF}}$. 
The test system is 4H-SiC supercell with the number of atoms of 576--4704. 
In both (a) and (b), data points obtained by KSDFT and OFDFT are plotted in red and blue colors, respectively. 
Each set of data points has been fitted by $AN^{\beta}$ with parameters $A$ and $\beta$, where $N$ is the number of atoms, 
and the fitted results are plotted as black solid lines. 
In (a), these fittings scale as $\mathcal{O}(N^{\beta})$ where $\beta=2.4,1.9,1.2,1.2,0.5$ from the top line to the bottom line.
$t_{1}$, $t_{2}$, and $t_{3}$ are plotted as circles, squares, and triangles, respectively.}
\label{comp_time} 
\end{figure}

Figure \ref{comp_time} (a) shows the results. The predominant computational time during a single SCF iteration in KSDFT is $t_{1}$, whereas it is $t_{2}$ or $t_3$ in OFDFT. The latter scales with $\mathcal{O}(N^{1.2})$ in our numerical confirmation. The overall computational time in our OFDFT scheme and in our KS-scheme using RSDFT code \cite{RSDFT1,RSDFT2,RSDFT3}, which may be the fastest available code, is shown in Fig.~\ref{comp_time}(b), indicating that the present OFDFT scheme has now achieved essentially the $\mathcal{O}(N)$ scaling.

Several $\mathcal{O}(N)$ density-functional calculations have been proposed and developed in the past. One primitive way is to introduce localized-orbital basis sets to express Kohn-Sham equations and then truncate the overlap of localized orbitals in the actual computations \cite{siesta,openmx}. This is obviously not based on a legitimate principle but relies on the incompleteness of the basis set practically. Another scheme is based on the ``nearsightedness principle"\cite{kohn-prl96} of many-electron systems, which states that principal quantities to describe physical properties are essentially local. One of such quantities may be the density matrix and the $\mathcal{O}(N)$ scheme with the density matrix has been developed\cite{conquest}. Yet in the actual computations, one need to truncate the density matrix in real space, which is actually a system dependent procedure. On the other hand, OFDFT does not require such system-dependent procedure once the suitable kinetic energy functional is developed. In this sense, OFDFT has a potential to become the most legitimate and practical  $\mathcal{O}(N)$ scheme, which is applicable to a broad range of materials on the equal footing.

\section{Conclusion}\label{sec:conc}

We have developed a scheme of the orbital-free density-functional-theory (OFDFT) calculations based on the accurate and transferrable kinetic-energy density functional (KEDF) which is created in an unprecedented way using appropriately constructed neural network (NN). Our OFDFT scheme has reproduced the electron density obtained in the state-of-the-art DFT calculations and then provided accurate structural properties of 24 different systems, ranging from atoms, molecules, metals, semiconductors and an ionic material. The accuracy and the transferability of our KEDF have been achieved by our NN training system in which the kinetic-energy functional derivative (KEFD) at each real-space grid point in diamond-structured carbon is used as the training data. The choice of the KEFD as the training data is essential in the following sense: First, it appears directly in the Euler equation which one should solve and it allows us a transparent and intuitive understanding, where the density and its derivatives are the primary and fundamental quantities in the spirit of DFT. Second, its leaning assists in reproducing the physical quantity expressed as the first derivative of the total energy. More generally, the present development of KEDF $T[\rho]$ is in the line of systematic expansion in terms of the functional derivatives $\delta^{\ell_1} T/\delta \rho^{\ell_1}$ through progressive increase of $\ell_1$. The present numerical success has demonstrated the validity of this approach with the detailed results for case of $\ell_1=1$. The computational cost of the present OFDFT scheme for the system size $N$ has indeed shown the scaling of $\mathcal{O}(N)$ inherent to OFDFT, as is evidenced by the computations of SiC consisting of thousands of Si and C atoms which are important in developing power-electronics devices.

\begin{acknowledgments}
This work was partly supported by the projects conducted under MEXT Japan named as 
``Priority Issue on Post-K computer" and ``Program for Promoting Research on the Supercomputer Fugaku". In the latter, we are involved in the two subprojects, ``Basic science for emergence and functionality in quantum matter: innovative strongly-correlated electron science by integration of Fugaku and frontier experiments" and ``Multiscale simulations based on quantum theory toward the developments of energy-saving next-generation semiconductor devices". The JSPS grants-in-aid (Grant Nos. 16H06345 and 18H03873) also support the present work. Computations were performed with the resources provided by HPCI System (Project ID: hp180170, hp180226, hp190145, hp190172, hp200122 and hp200132) and 
by Supercomputer Center at Institute for Solid State Physics, University of Tokyo and at Institute for Molecular Sciences, Natural Institute of Natural Sciences 
\end{acknowledgments}

\appendix

\section{Construction of {\it ab initio} Local IONIC Pseudopotentials}\label{append:LIPS}

In this appendix, we explain our scheme to generate {\it ab-initio} local ionic pseudopotentials (LIPSs), its application to 7 elements, lithium, carbon, sodium, aluminum, silicon, chlorine, and copper, and the validity and the transferability of the generated NLPSs. 

Local pseudopotentials combined with OFDFT are developed in the pioneering work 
by Carter and her collaborators \cite{zhou-wang-carter-prb04,huang-carter-pccp08,huang-carter-prb10,delrio-dieterich-carter-jctc17}. They are called ``bulk-derived local pseudopotential" (BLPS) since the potentials contain parameters which are optimized to reproduce structural properties of target bulk materials: The parameters are 
(1) the value of the non-Coulombic part of the BLPS in reciprocal space at $G=0$ and (2) the cutoff radius $r_{\rm c}$ beyond which the Coulombic tail is imposed on the BLPS in real space. In the construction of BLPS, the KSDFT calculations with NLPS are first performed to obtain the bulk modulus $B_0$, equilibrium volumes $V_0$, and also the energy ordering of various phases. Then the above parameters in BLPS are adjusted to reproduce those structural characteristics obtained by the KSDFT calculations. Such fitting is also improved by minimizing the difference between forces obtained by the BLPS with OFDFT and those by an NLPS with KSDFT \cite{delrio-dieterich-carter-jctc17}. Currently, these BLPSs are available for Al, As, Ga, In, Li, Mg, P, Sb, Si \cite{blps-github}. After the appropriate fitting, the bulk properties calculated from the BLPSs are in good agreement with those by NLPSs \cite{huang-carter-pccp08}. 

In spite of relatively good performance of BLPSs, the above mentioned fitting process is manual and cumbersome. 
The iterative improvement of the BLPS is unavoidable until required target properties can be obtained within acceptably small errors. Furthermore, the previous procedure to solve KS equation inversely and obtain the effective potential \cite{zhou-wang-carter-prb04,huang-carter-pccp08,huang-carter-prb10} using the Wang-Parr method\cite{wang-parr-pra93} or the Wu-Yang method\cite{wu-yang-jcp03} require good initial guess for the effective potential.

In our work, we aim to eliminate the fitting parameters and to circumvent the iterative fitting procedure.
In our approach, the LIPS 
can be treated solely in reciprocal space 
and thus there is no need to perform Fourier-Bessel transform which requires elaborate interpolation in reciprocal space. 

The first step is to find an effective potential $V_{\rm eff}({\bm r})$ that reproduces a target electron density generated by 
the KSDFT calculations using NLPSs. 
Among many ways to invert the KS equation, we adopt the Kadantsev-Stott method \cite{kadantsev-stott-pra04} based on the Haydock-Foulkes variational principle \cite{haydock-foulkes-prb89}, 
which does not require a particularly good initial guess for the local pseudopotential. 
In this method, a functional $\Upsilon[V_{\rm eff}({\bm r})]$ is defined as
\begin{equation}\label{upsilon}
 \Upsilon[V_{\rm eff}({\bm r})]=-\sum_{i:{\rm occ}}\epsilon_{i}[V_{\rm eff}({\bm r})] + \int V_{\rm eff}({\bm r}) \rho_{\rm target}({\bm r}) d{\bm r},
\end{equation}
where $\epsilon_{i}$ are eigenvalues corresponding to the KS orbitals $\{\phi_{i}\}$, and $\rho_{\rm target}({\bm r})$ is the target electron density obtained by 
the KSDFT calculation with NLPSs, which is 
to be reproduced here. The Haydock-Foulkes variational principle ensures that the true effective potential corresponding to the target density satisfies the stationary condition ${\delta\Upsilon[V_{\rm eff}]}/{\delta V_{\rm eff}}=0$.

In the $k$-th iteration of the minimization of $\Upsilon[V]$, the KS equation is solved inversely using the $V_{\rm eff}^{(k-1)}({\bm r})$ in the previous iteration to obtain the density $\rho^{(k)} (\bm r)$. 
We note that the functional derivative of $\Upsilon$ satisfies an identity ${\delta \Upsilon[V]} / {\delta V} = - \rho({\bm r}) + \rho_{\rm target}({\bm r})$. Hence in the $k$-th iteration, $V_{\rm eff}^{(k)}({\bm r})$ is explored along a line defined by 
\begin{equation}
 V_{\rm eff}^{(k)}({\bm r}) = V_{\rm eff}^{(k-1)}({\bm r}) 
   - \alpha_{1} 
\left. \frac{\delta \Upsilon [V]}{\delta V}\right|_{\rho=\rho^{(k)}} ,
\end{equation}
and $\alpha_1$ is determined by the line search. 
We iteratively continue this process until $\sigma$,
\begin{equation}\label{def-sigma}
 \sigma = \sqrt{ \frac{1}{\Omega} \int_{\Omega} |\rho^{(k)}({\bm r})-\rho_{\rm target}({\bm r})|^2 d{\bm r}},
\end{equation}
becomes less than the preset value. We have implemented this variational minimization method in the RSDFT code\cite{RSDFT1,RSDFT2,RSDFT3}.

In actual computations, we first generate the target bulk electron densities 
by solving the KS equations using NLPSs. 
We then invert the KS equation using these target bulk electron densities to obtain the local effective potentials $V_{\rm eff}({\bm r})$ for these targets using the Kadantsev-Stott method explained above. 
The settings of $k$-point grids and real-space grid spacing used 
in the inversion are the same as those used in generating the target bulk electron densities.

Next, the Hartree potential $V_{\rm H}({\bm r})$ and exchange-correlation potential $V_{\rm xc}(\bm r)$ 
among valence electrons are subtracted from $V_{\rm eff}({\bm r})$ to extract the local ionic potential for the bulk: 
\begin{equation}\label{vloc-veff}
 V_{\rm bulk}^{\rm ion} ({\bm r})=V_{\rm eff}({\bm r}) - V_{\rm H} ({\bm r}) - V_{\rm xc}(\bm r) . 
\end{equation}
We then convert $V_{\rm bulk}^{\rm ion} ({\bm r})$ into the atom-centered local 
ionic pseudopotential in reciprocal space $\tilde{V}_{\rm loc}^{\rm ion}(G)$ as follows. 
First, $V_{\rm bulk}^{\rm ion} ({\bm r})$ is Fourier-transformed into reciprocal space as 
\begin{equation}
V_{\rm bulk}^{\rm ion} ({\bm G}) = \frac{1}{\Omega} \int_{\Omega} V_{\rm bulk}^{\rm ion} ({\bm r}) e^{ i {\bm G} \cdot {\bm r}} d{\bm r},
\end{equation}
where $\Omega$ is the unit cell volume. Then using the structure factor $S ( \bm G )$ 
of the target bulk material, 
we obtain the Fourier transformed atom-centered ionic pseudopotential: 
\begin{equation}
V^{\rm ion} ({\bm G}) = \frac{V_{\rm bulk}^{\rm ion} ({\bm G}) }{S({\bm G})}. 
\end{equation}
Finally, by spherically averaging $V^{\rm ion} ({\bm G}) $, we obtain the atom-centered local ionic pseudopotential in reciprocal space, $\tilde{V}_{\rm loc}^{\rm ion}(G)$: 
\begin{equation}\label{Vloc_atom_G}
 \tilde{V}_{\rm loc}^{\rm ion}(G) = \frac{1}{N_G} \sum_{|{\bm G}_i|=G} V^{\rm ion} ({\bm G}_i) ,
\end{equation}
where $N_G$ is the number of $G$ vectors with the length being $G$. The procedure to construct $\tilde{V}_{\rm loc}^{\rm ion}(G)$ explained above ensures that the resulting $\tilde{V}_{\rm loc}^{\rm ion}$ has the proper Coulombic tail due to the nucleus and core electrons 
in real space. Hence we fit $\tilde{V}_{\rm loc}^{\rm ion}(G)$ numerically obtained above to the form, 
\begin{eqnarray}
 \tilde{V}_{\rm loc}^{\rm ion}(G) &=& \frac{4\pi}{\Omega} \int_{0}^{\infty} r^{2} j_{0}(Gr) \tilde{V}_{\rm loc}^{\rm ion}(r) dr \nonumber \\
 &=& -\frac{4\pi Z}{\Omega G^2} \sum_{i=1}^{2}c_{i}^{\rm core}\exp{ \left[ -\frac{G^2}{4\alpha_{i}^{\rm core}} \right] } \nonumber \\
 &+& \frac{\sqrt{(2\pi)^3}}{\Omega}{r_c}^{3} \exp{ \left[ -\frac{(Gr_c)^2}{2} \right] } \nonumber\\
 &\times&\left\{ C_1 + C_2[3-(Gr_c)^2] \right. \nonumber \\
 &+&C_3[15-10(Gr_c)^2+(Gr_c)^4]  \nonumber \\
 &+& \left. C_4[105-105(Gr_c)^2+21(Gr_c)^4-(Gr_c)^6] \right. \nonumber \\
 &+& \left. C_5[945-1260(Gr_c)^2+378(Gr_c)^4 \right. \nonumber \\
 &-& \left. 36(Gr_c)^6+(Gr_c)^8] \right\}, 
\end{eqnarray}
which corresponds to the Fourier-inversed-transform, 
\begin{eqnarray}
 \tilde{V}_{\rm loc}^{\rm ion}(r) &=& -\frac{Z}{r} \sum_{i=1}^{2} c_{i}^{\rm core}
 {\rm erf}\left[(\alpha_{i}^{\rm core})^{1/2} r\right] \nonumber\\
 &+&\exp{ \left[ -\frac{1}{2} \left(\frac{r}{r_c}\right)^2 \right] } \sum_{n=1}^{5} C_{n} \left(\frac{r}{r_c}\right)^{2n-2},
\end{eqnarray}
where $c_{1}^{\rm core}+c_{2}^{\rm core}=1$ \cite{bhs-82,gth-96}.

The KSDFT calculations to obtain the target bulk densities were performed using the RSDFT code\cite{RSDFT1,RSDFT2,RSDFT3}. We adopted the generalized gradient approximation of Perdew, Burke, and Ernzerhof\cite{perdew-burke-ernzerhof-prl96} for the exchange-correlation functional. We used Troullier-Martins (TM) NLPSs\cite{troullier-martins-prb91-1,troullier-martins-prb91-2}. As target densities, we chose the densities of ds-Si, ds-C, bcc-Li, bcc-Na, fcc-Al, fcc-Cu, and NaCl with the lattice constants of 5.466 {\AA}, 3.560 {\AA}, 3.430 {\AA}, 4.212 {\AA}, 4.051 {\AA}, 3.634 {\AA}, and 5.694 {\AA}, respectively, all of which are equilibrium values within the KSDFT scheme using NLPSs. The $k$-point sampling grids were taken to be $4\times4\times4$ in all calculations. The real-space grid spacing was 0.190 {\AA} for Si, 0.178 {\AA} for C, 0.172 {\AA} for Li, 0.211 {\AA} for Na, 0.169 {\AA} for Al, 0.151 {\AA} for Cu, and 0.190 {\AA} for NaCl.

The convergence criteria of the KS inversion was set to be 
$\sigma<3.5\times10^{-4}$ ${\rm bohr}^{-3}$ for Li, $\sigma<1.0\times10^{-3}$ ${\rm bohr}^{-3}$ for C, 
$\sigma<2.0\times10^{-5}$ ${\rm bohr}^{-3}$ for Na, $\sigma<1.0\times10^{-5}$ ${\rm bohr}^{-3}$ for Al, 
$\sigma<1.0\times10^{-4}$ ${\rm bohr}^{-3}$ for Si, $\sigma<3.0\times10^{-5}$ ${\rm bohr}^{-3}$ for NaCl, and 
$\sigma<8.5\times10^{-3}$ ${\rm bohr}^{-3}$ for Cu.
We minimized the functional $\Upsilon[V_{\rm eff}({\bm r})]$ via the conjugate gradient (CG) method. 
The obtained parameters for  Li, C, Na, Al, Si, Cl, and Cu LIPSs are shown in Table \ref{tab:lps_param}. 

\begin{table*}[!h]
 \begin{center}
 \caption{Parameters of local ionic pseudopotentials for Li, C, Na, Al, Si, Cl, and Cu.}
 \begin{tabular}{l ccccccccc} \hline\hline
    &   $c_{1}^{\rm core}$ & $\alpha_{1}^{\rm core}$ & $\alpha_{2}^{\rm core}$ & $r_{c}$ & $C_{1}$ & $C_{2}$ & $C_{3}$ & $C_{4}$ & $C_{5}$ \\ \hline
 Li &  1.0000 & 3.1250 & --      & 0.4000 & -3.12247 & -5.29585 & 1.29259 & -0.0299128 & 0  \\
 C  &  1.0000 & 1.4635 & --  & 0.5845 & 8.4107 & -13.007 & 4.8809 & -0.6743  & 0.02793 \\
 Na  &  1.0000 & 1.1619 & --    & 0.6560 & -3.85643 & -8.09377 & 2.90894 & -0.190011  & 0 \\
 Al &  1.0000 & 0.5732 & -- & 0.9340 & 3.22841 & -1.4132 & 0.147102 & -0.00494713 & 0  \\
 Si &  1.6054 & 2.1600 & 0.8600 & 0.7999 & 9.0231 & -3.7692 & 0.5453 & -0.02952 & 0  \\
 Cl &  1.0000 & 1.3171 & -- & 0.616128& 5.29287 & -2.12203 & 0.169072 & -0.014369& 0  \\
 Cu  &  1.0000 & 1.5097 & --     & 0.5755 & -2.58512 & -17.0765 & 5.29496 & -0.31904 & 0 \\ \hline\hline
 \end{tabular}
 \label{tab:lps_param}
 \end{center}
\end{table*}

\begin{figure*}[!h]
\centering
\includegraphics[width=.8\linewidth]{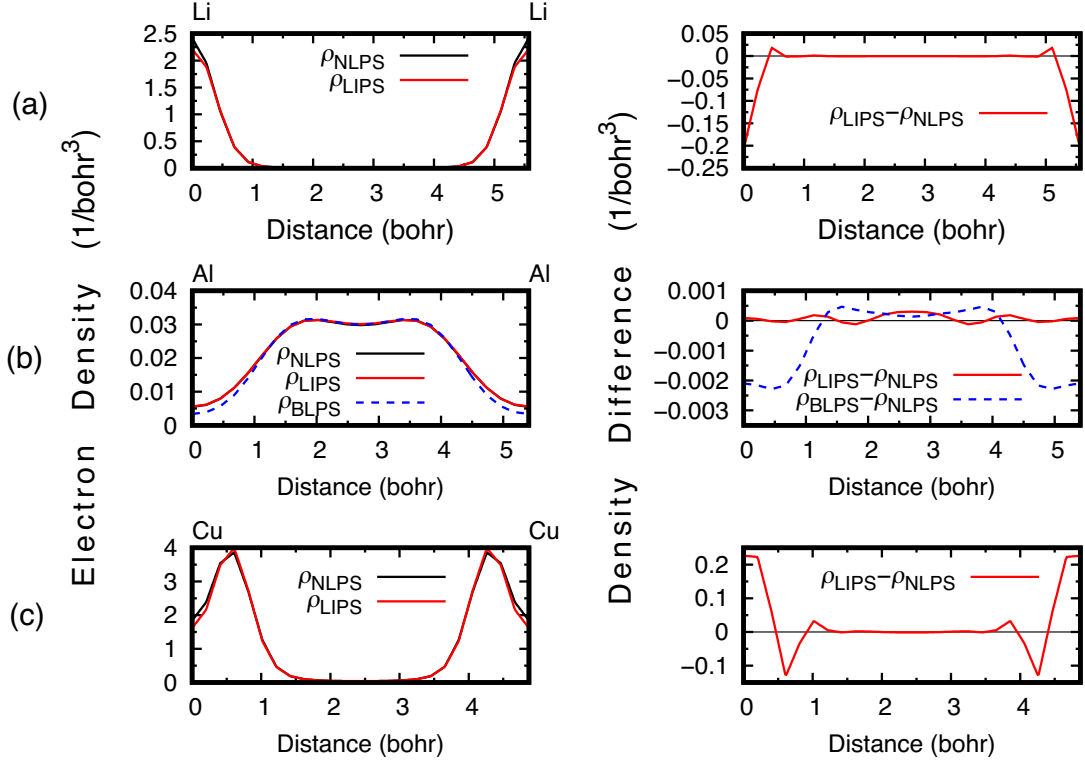}
\caption{
Electron densities (left panels) obtained by NLPS and LIPS and the electron-density differences (right panels) for (a) body-centered-cubic(bcc)-Li, 
(b) face-centered-cubic(fcc)-Al, and (c) fcc-Cu along the [100] direction. The position of each atom along the [100] is marked on the top lines of each figure. For Al, the data obtained by using BLPS is also shown.}
\label{rho_metal_nlps_vs_lps}
\end{figure*}
%
\begin{figure*}[!h]
\centering
\includegraphics[width=.8\linewidth]{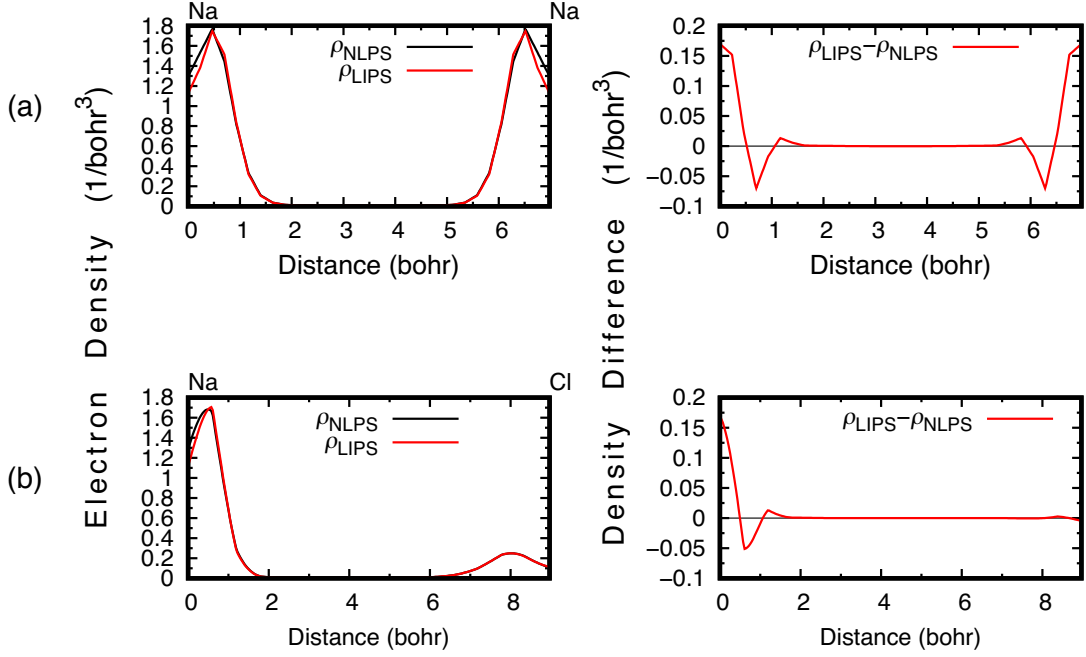}
\caption{
Electron densities (left panels) obtained by NLPS and LIPS and the electron-density differences (right panels) for (a) body-centered-cubic(bcc)-Na along the [100] direction 
and (b) NaCl along the [111] direction on the (110) plane. The position of each atom along each direction is marked on the top lines of each figure.}
\label{rho_nacl_nlps_vs_lps}
\end{figure*}
%
\begin{figure*}[!h]
\centering
\includegraphics[width=.8\linewidth]{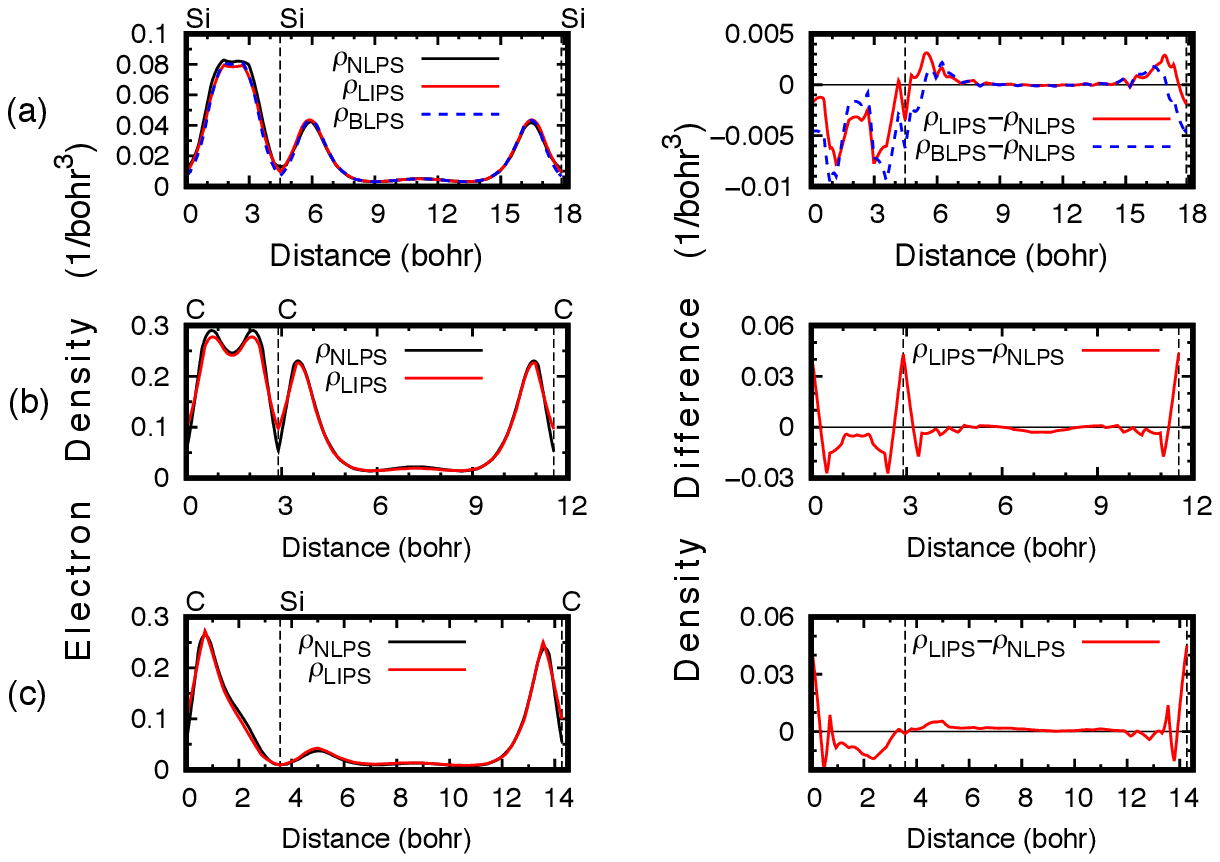}
\caption{
Electron densities (left panels) obtained by NLPS and LIPS and the electron-density differences (right panels) for (a) diamond-structured Si, (b) diamond-structured C, and 
(c) SiC along the [111] direction on the (110) plane. The position of each atom along the [111] on the (110) is marked on the top lines of each figure. For Si, the data obtained by using BLPS is also shown.}
\label{rho_sic_nlps_vs_lps}
\end{figure*}

We now assess the quality of our LIPSs. 
We compare the electron densities of bcc-Li, fcc-Al, fcc-Cu, bcc-Na, NaCl, ds-Si, ds-C, and zincblende(3C)-SiC calculated with the NLPS and our LIPS. 
We denote the density obtained by the NLPS as $\rho_{\rm NLPS}$ and that by our LIPS as $\rho_{\rm LIPS}$. 
Figures~\ref{rho_metal_nlps_vs_lps},\ref{rho_nacl_nlps_vs_lps},and \ref{rho_sic_nlps_vs_lps} show the electron densities of the target materials along selected directions: 
bcc-Li along $[100]$, fcc-Al along $[100]$, fcc-Cu along $[100]$, bcc-Na along $[100]$, NaCl along $[111]$, ds-Si along $[111]$, ds-C along $[111]$, and SiC along $[111]$. 
For Al and Si, the data from BLPS is also available. 
Hence we plot the density obtained by the BLPS, labelled as $\rho_{\rm BLPS}$ in Figs. \ref{rho_metal_nlps_vs_lps} and \ref{rho_sic_nlps_vs_lps}. 
Figures \ref{rho_metal_nlps_vs_lps}, \ref{rho_nacl_nlps_vs_lps}, and \ref{rho_sic_nlps_vs_lps} clearly show that the electron density obtained by our LIPS well reproduces the density by NLPS. The density by BLPS also seems to be good enough. For Al [Fig.~\ref{rho_metal_nlps_vs_lps} (b)] and Si [Fig.~\ref{rho_sic_nlps_vs_lps} (a)], it appears that $|\rho_{\rm LIPS} - \rho_{\rm NLPS} |$ is generally smaller than $| \rho_{\rm BLPS} - \rho_{\rm NLPS} |$. As for diamond C and SiC, the calculated $\rho_{\rm LIPS}$ satisfactorily reproduce the characteristic features of $\rho_{\rm NLPS}$: The peculiar double peaks between the C-C bonds in diamond C and the substantial iconicity in SiC (Fig.~\ref{rho_sic_nlps_vs_lps}). The maximum deviation of $\rho_{\rm LIPS} - \rho_{\rm NLPS}$ for diamond is $ 4.38 \times10^{-2}$ ${\rm bohr}^{-3}$ at the C nuclear site for diamond C. For SiC it is $4.49 \times 10^{-2}$ ${\rm bohr}^{-3}$ again at the C nuclear site. 

We also evaluate the quality of our LIPSs by calculating the total energy as a function of the volume $E(V)$ for 
ds-Si, ds-C, bcc-Li, bcc-Na, fcc-Al, fcc-Cu, NaCl, and zincblende(3C)-SiC. The obtained $E(V)$ is fitted to Murnaghan's equation of state \cite{murnaghan_eqs}
to deduce the equilibrium lattice constant $a_0$ and the bulk modulus $B_0$. 
The results are shown in Table \ref{tab:a0_B0} along with those obtained by the NLPSs. The results from the BLPS for Li, Al, and Si are also tabulated. 

It is clear that the structural properties of the 8 different materials produced by our LIPSs are as accurate as those from the NLPSs. 
The electron densities and the structural properties obtained above certainly assures the reliability and the transferability of the frozen-core approximation (pseudopotential scheme) using our LIPSs.

\begin{table*}
 \begin{center}
 \caption{Equilibrium lattice constants ($a_{0}$) in {\AA} and bulk moduli ($B_{0}$) in GPa 
 obtained by using the NLPS and our LIPS 
of 8 different solids. For bcc-Li, fcc-Al and ds-Si, the obtained values using the BLPS are also listed. 
Numbers in parentheses are the relative errors with respect to the experimental values. The mean absolute relative errors (MAREs) in \% with respect to the experimental values are 0.59 (NLPS), 1.11 (BLPS), and 1.13 (LIPS) for $a_{0}$, and 6.21 (NLPS), 6.78 (BLPS), 11.0 (LIPS) for $B_{0}$.}
\begin{tabular}{c|cc cc cc cc cc cc cc cc} \hline\hline
                                      & \multicolumn{2}{c}{bcc-Li} & \multicolumn{2}{c}{diamond} & \multicolumn{2}{c}{bcc-Na} & \multicolumn{2}{c}{fcc-Al} & \multicolumn{2}{c}{ds-Si} & \multicolumn{2}{c}{NaCl} & \multicolumn{2}{c}{fcc-Cu} & \multicolumn{2}{c}{SiC} \\ 
                                      & $a_{0}$ & $B_{0}$ & $a_{0}$ & $B_{0}$ & $a_{0}$ & $B_{0}$  & $a_{0}$ & $B_{0}$ & $a_{0}$ & $B_{0}$ & $a_{0}$ & $B_{0}$ & $a_{0}$ & $B_{0}$ &  $a_{0}$ & $B_{0}$ \\ \hline
 \multirow{2}{*}{NLPS}     &  3.430 & 13.3 & 3.560 & 434 & 4.212 & 7.68 & 4.051 & 77.8 & 5.466 & 86.9 & 5.694 & 23.9 & 3.634 & 146  & 3.077  & 226 \\
                                     &  (-1.75) & (-4.32) & (-0.20) & (-1.81) & (-0.31) & (21.9) & (0.02) & (2.37) & (0.63) & (-12.0) & (0.96) & (-2.05)  & (0.66) & (4.29)   & (-0.19) & (0.89) \\
 \multirow{2}{*}{BLPS}  & 3.481 & 14.8  & -- & --  & --  & --  & 3.968 & 84.0    & 5.377 & 95.5 &  --  & --  &  --  & --  &  --   &  --   \\
                        & (-0.29) & (6.47) &    &  &  &        & (-2.02) & (-9.08) & (-1.01)  & (-17.1) & &     &      &     &   & \\
 \multirow{2}{*}{LIPS}   & 3.508 & 14.9  & 3.517 & 327   & 4.266  & 7.71   & 4.050            & 77.5    & 5.392   & 100 & 5.480 & 25.9 & 3.645         & 158 & 3.023 & 246  \\
                         & (0.49) & (7.19)   & (-1.40) & (-26.0) & (0.97) & (22.4) & (0.00) & (1.97) & (-0.74) & (1.21) & (-2.84) & (6.15) & (0.97)  & (12.9) & (-1.62) & (9.82)  \\ \hline
Exp.       & 3.491 & 13.9   & 3.567 & 442   & 4.225 & 6.3  & 4.05 & 76    & 5.432   & 98.8 & 5.64 & 24.4 & 3.61 & 140 & 3.083 & 224  \\ \hline\hline
 \end{tabular}
 \label{tab:a0_B0}
 \end{center}
\end{table*}

\section{Functional Derivative of Laplacian-level Kinetic Energy Functional}\label{append:KEFD_formula}

For any Laplacian level KEDF of the form
\begin{equation}
 T[\rho]=\int \tau(\rho,|\nabla\rho|^{2},\nabla^{2}\rho)d{\bm r}, 
\end{equation}
the variation with respect to the electron density is
\begin{eqnarray}\label{func_deriv1}
 \delta T[\rho]&=&\int \left[ \frac{\partial\tau}{\partial\rho}\delta\rho + \frac{\partial\tau}{\partial(|\nabla\rho|^{2})} \delta(|\nabla\rho|^{2}) \right. \nonumber \\
        &&+ \left. \frac{\partial\tau}{\partial(\nabla^{2}\rho)} \delta(\nabla^{2}\rho) \right] d{\bm r} \nonumber \\
         &=&\int \left[ \frac{\partial\tau}{\partial\rho} - 2\nabla\cdot\left(\frac{\partial\tau}{\partial(|\nabla\rho|^{2})}\nabla\rho\right) \right. \nonumber \\ 
         &&+ \left. \nabla^{2}\left(\frac{\partial\tau}{\partial(\nabla^{2}\rho)}\right) \right] \delta\rho d{\bm r}, 
\end{eqnarray}
where we used $\delta(|\nabla\rho|^{2})=2\nabla\rho\cdot\nabla(\delta\rho)$, $\delta(\nabla^{2}\rho)=\nabla^{2}(\delta\rho)$, 
and the first ($\int{\bm v}({\bm r})\cdot\nabla f({\bm r)}d{\bm r}=-\int f({\bm r})\nabla\cdot v({\bm r})d{\bm r}$) 
and the second Green's identities ($\int f({\bm r})\nabla^{2}g({\bm r})d{\bm r}=\int \nabla^{2}f({\bm r})g({\bm r})d{\bm r}$). 
Specifically, when the kinetic energy is expressed in meta-GGA form (Eq.~(\ref{NN_enhance_fac})), substituting the expressions 
\begin{equation}
 \frac{\partial\tau}{\partial\rho}=\tau^{\rm TF} \left( \frac{5}{3\rho}F+\frac{\partial s^{2}}{\partial \rho}\frac{\partial F}{\partial s^{2}}
                                  + \frac{\partial q}{\partial \rho}\frac{\partial F}{\partial q} \right)
\end{equation}
\begin{equation}
 \frac{\partial\tau}{\partial(|\nabla\rho|^{2})}=\tau^{\rm TF}\frac{\partial F}{\partial s^{2}} \frac{\partial s^{2}}{\partial(|\nabla\rho|^{2})}
\end{equation}
\begin{equation}
 \frac{\partial\tau}{\partial(\nabla^{2}\rho)}=\tau^{\rm TF}\frac{\partial F}{\partial q} \frac{\partial q}{\partial(\nabla^{2}\rho)}
\end{equation}
in Eq.~(\ref{func_deriv1}) leads to Eq.~(\ref{def_KEFD_NN}) 
with the use of ${\partial s^{2}}/{\partial \rho}=-{8s^{2}}/(3{\rho})$, ${\partial s^{2}}/{\partial(|\nabla\rho|^{2})} = {\rho^{-8/3}}/{(4(3\pi^2)^{2/3})}$, 
${\partial q}/{\partial \rho}=-{5q}/(3{\rho})$, and  ${\partial q}/{\partial(\nabla^{2}\rho)} = {\rho^{-5/3}}/{(4(3\pi^2)^{2/3})}$.

\section{Training details}\label{append:train_detail}
We here determine the training hyperparameters, $\eta$ and $\nu$, appearing in the NN weight updating formula Eq.~(\ref{WIter}). 
We have first performed a grid search for optimum constant values of $\eta$ and $\nu$ in the range of $ 10^{-4} \le \eta \le 10$ and $ 10^{-8} \le \nu \le 1$, and 
found that a pair of constants, ($\eta = 0.1, \nu = 10^{-5}$), provides the smallest RMSE $\sqrt{2L}$ of 0.228. 
Since the ordinary gradient scheme represented by $\nu$ is helpful only at early stages of training, we introduce proper time scheduling of $\nu$, keeping $\eta=0.1$ constant. We consider three options for $\nu (t)$, $\nu (t) = \nu_{0}/(1+bt)$, $\nu_{0}/\sqrt{1+bt}$, and $\nu_{0}\exp(-bt)$, 
where $t$ is the number of the training epoch and $\nu_{0}=10^{-5}$. 
We have found that 
$\nu=\nu_{0}/(1+bt)$ with $b=10^{-2}$ 
provides the smallest RMSE and adopted this scheduling 
in the present work.

\begin{figure*}
\begin{center}
\includegraphics[width=.6\linewidth]{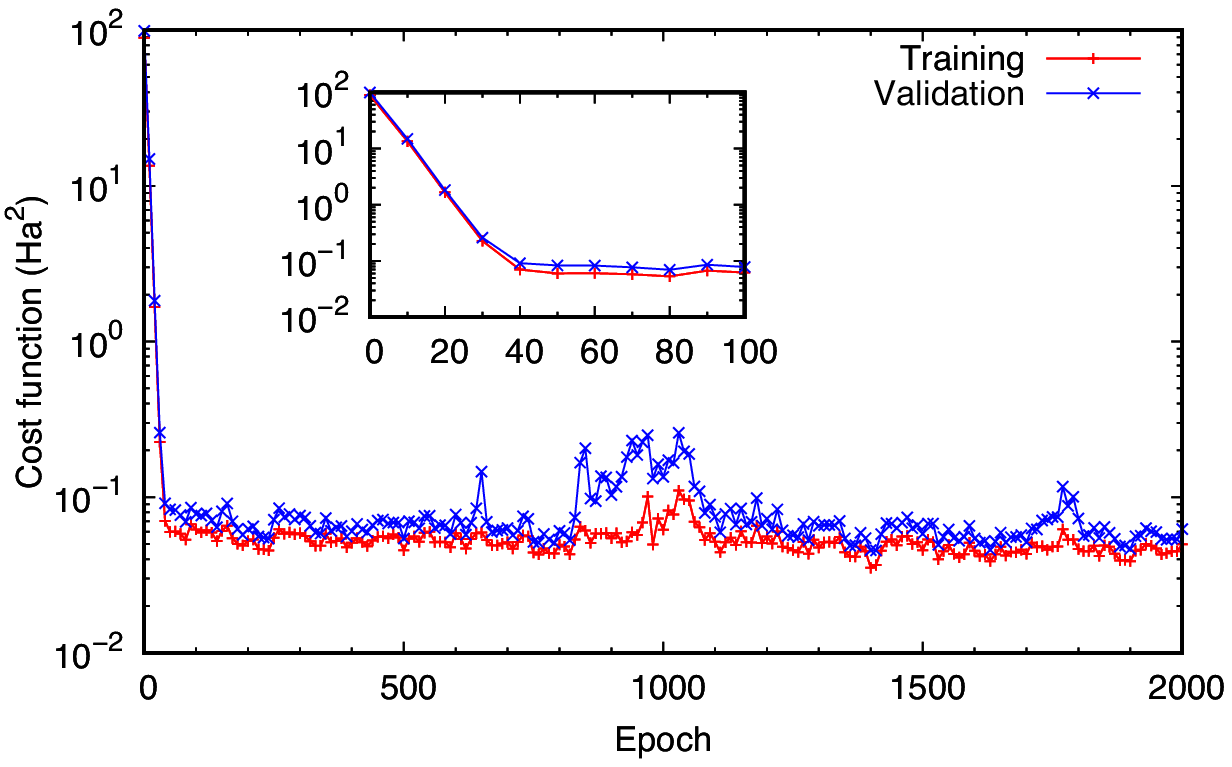}
\caption{Evolution of the cost function during the optimization process. The inset shows a zoom-in for the epochs smaller than 100.}\label{cost_df}
\end{center}
\end{figure*}

Figure~\ref{cost_df} shows variation of the cost function $L$ during the optimization
for the case of the fixed $A = 10^{1.5}$ (see Appendix \ref{append:A_opt} below). 
It is noteworthy that the training of only 100 epochs makes the cost function small by three-orders-of-magnitude, being less than 0.1 Ha$^{2}$. 
It is shown that the cost function becomes smallest at the 1400-th epoch. We thus adopt the NN weight at the 1400-th epoch as the final trained weight. The total computational time for the optimization of the NN weights is typically less than 30 minutes on usual house computers, which is the orders-of-magnitudes shorter than the computational time for the SCF calculations. 
We note that 3000 training points are sufficient to minimize the cost function to the value less than 0.06 Ha$^{2}$. 

\section{Functional derivative training}\label{append:KEFD_train}

In order to calculate $\partial L/\partial {\bm W}$, we need to obtain several derivatives, ${\partial F^{\text{NN}}}/{\partial {\bm W}}$, $\partial({\partial F^{\text{NN}}}/{\partial(s^{2})})/\partial {\bm W}$ and $\partial({\partial F^{\text{NN}}}/{\partial q})\partial {\bm W}$. In this work, the first derivative is obtained by the conventional backpropagation, and the second and the third derivatives are obtained by the backpropagation for the derivatives \cite{pukritt-neural_netw11,pukritt-jchemphys09}. Here we briefly overview the algorithm of the backpropagation in order to explain how to compute the partial derivative of NN outputs with respect to the NN weights $\bm W$. The number of the neuron in the final layer is  $D_N = 1$ in our case. 
In the conventional backpropagation, our aim is to compute ${\partial z_{1}^{(N)}}/{\partial W_{ji}^{(l)}}$ 
from ${\partial z_{1}^{(N)}}/{\partial a_{1}^{(N)}}={\sigma^{(N)}}^{\prime}(a_{1}^{(N)})$. 
Using the quantity $\delta_{j}^{(l)}$ that satisfies the recursive formula 
\begin{eqnarray}
 \delta_{j}^{(l)} &\equiv& \frac{\partial z_{1}^{(N)}}{\partial a_{j}^{(l)}}
 =\sum_{k} \frac{\partial z_{1}^{(N)}}{\partial a_{k}^{(l+1)}} \frac{\partial a_{k}^{(l+1)}}{\partial a_{j}^{(l)}} \nonumber \\ 
 &=&\sum_{k} \delta_{k}^{(l+1)} \frac{\partial}{\partial a_{j}^{(l)}} \sum_{i}W_{ki}^{(l+1)}\sigma^{(l)}(a_{i}^{(l)}) \nonumber \\
 &=&\sum_{k} \delta_{k}^{(l+1)} W_{kj}^{(l+1)} {\sigma^{(l)}}^{\prime}(a_{j}^{(l)}), 
\end{eqnarray}
we can obtain ${\partial z_{1}^{(N)}}/{\partial W_{ji}^{(l)}}$ as 
\begin{equation}\label{backprop1}
 \frac{\partial z_{1}^{(N)}}{\partial W_{ji}^{(l)}}=\delta_{j}^{(l)}z_{i}^{(l-1)}. 
\end{equation}
This formula allows us to recursively compute ${\partial z_{1}^{(N)}}/{\partial W_{ji}^{(l)}}$ 
from $\delta_{1}^{(N)}$. 
Next, we derive a similar recursive method to obtain $\partial \gamma_{1r}^{(N)}/\partial W_{ji}^{(l)}$, 
where $\gamma_{1r}^{(N)}=\partial z_{1}^{(N)}/\partial z_{r}^{(1)}$ is the derivative of the output with respect to the inputs. 
Here $\gamma_{1r}^{(l)}$ has been computed recursively as
\begin{equation}
 \gamma_{1r}^{(l)}={\sigma^{(l)}}^{\prime}(a_{j}^{(l)})\sum_{k}W_{jk}^{(l)}\gamma_{kr}^{(l-1)} 
\end{equation}
in advance. 
Using Eq.~(\ref{backprop1}), we have 
\begin{eqnarray}
 \frac{\partial \gamma_{1r}^{(N)}}{\partial W_{ji}^{(l)}}
 &=&z_{i}^{(l-1)}\frac{\partial \delta_{j}^{(l)}}{\partial z_{r}^{(1)}} + \frac{\partial z_{i}^{(l-1)}}{\partial z_{r}^{(1)}}\delta_{j}^{(l)} \nonumber\\
 &=&z_{i}^{(l-1)}\frac{\partial \delta_{j}^{(l)}}{\partial z_{r}^{(1)}} + \gamma_{ir}^{(l-1)}\delta_{j}^{(l)}. 
\end{eqnarray}
By introducing a quantity $\zeta_{jr}^{(l)}$ that satisfies the recursive formula 
\begin{eqnarray}
 \zeta_{jr}^{(l)} &\equiv& \frac{\partial \delta_{j}^{(l)}}{\partial z_{r}^{(1)}}
 =\frac{\partial}{\partial z_{r}^{(1)}}\left[ {\sigma^{(l)}}^{\prime}(a_{j}^{(l)})\sum_{k}\frac{\partial z_{1}^{(N)}}{\partial a_{k}^{(l+1)}}W_{kj}^{(l+1)} \right] \nonumber \\
 &=&\frac{\partial {\sigma^{(l)}}^{\prime}(a_{j}^{(l)})}{\partial z_{j}^{(l)}}\gamma_{jr}^{(l)}\sum_{k}W_{kj}^{(l+1)}\delta_{k}^{(l+1)} \nonumber\\
 &+&{\sigma^{(l)}}^{\prime}(a_{j}^{(l)})\sum_{k}W_{kj}^{(l+1)}\zeta_{kr}^{(l+1)}, 
\end{eqnarray}
we can compute ${\partial \gamma_{1r}^{(N)}}/{\partial W_{ji}^{(l)}}$ starting from $\zeta_{jr}^{(N)}=0$ 
since we take the activation function in the output layer 
as an identity function, $z_{1}^{(N)}=\sigma^{(N)}(a_{1}^{(N)})=a_{1}^{(N)}$.

\section{Determination of the enhancement factor $ F^{(0)}(s^{2},q) $}\label{append:F0_fit}

We have optimized $\beta$ 
in the enhancement factor $ F^{(0)}(s^{2},q) $, Eq.~(\ref{def_PGSL}), so that 
the inverse of the response function derived from this enhancement factor in the homogeneous-gas limit, 
\begin{equation}\label{fitted_resp}
 -\frac{1}{\chi_0 (\eta)} = \frac{\pi^{2}}{k_{F}}\left( 1+\frac{\eta^{2}}{3}+\frac{9}{5}\beta\eta^{4} \right)
\end{equation}
reproduces the Lindhard function: 
\begin{equation}
 -\frac{1}{\chi_{\text{Lind}}(\eta)}=\frac{\pi^{2}}{k_{F}}\left( \frac{1}{2}+\frac{1-\eta^{2}}{4\eta}\ln \left| \frac{1+\eta}{1-\eta} \right|  \right)^{-1}. 
\end{equation}
Here $\eta = k / (2 k_{\rm F})$ with $k_{\rm F}$ being $(3 \pi^2 \rho)^{1/3}$.

\begin{figure}
\begin{center}
\includegraphics[width=.9\linewidth]{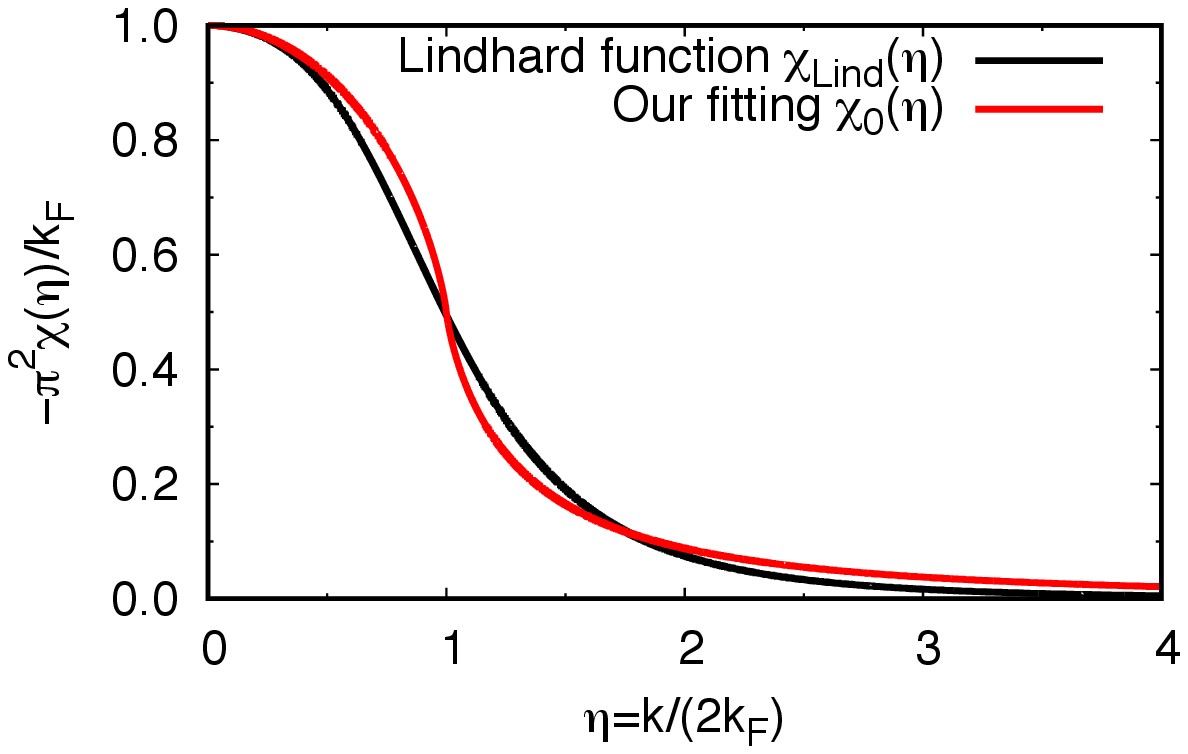}
\caption{Comparison of our response function $\chi_{0}(\eta)$ with the Lindhard response function $\chi_{\text{Lind}}(\eta)$ after the optimization of the parameter $\beta$.}
\label{linear_resp}
\end{center}
\end{figure}

To this end, we have discretized $\eta$ and defined $10^{4}$ points $\eta_{i}\in[10^{-5},4]$. 
Then, we have performed the least square fitting that minimizes 
the cost function $C$
\begin{equation}
 C = \frac{1}{2} \sum_{i} \left[ \frac{1}{\chi_0 (\eta_{i})}-\frac{1}{\chi_{\text{Lind}}(\eta_{i})} \right]^{2}, 
\end{equation}
which results in the optimized value, $\beta=0.382$. 
The obtained $\chi_0 (\eta)$ is compared with $\chi_{\text{Lind}}(\eta)$ in Fig.~\ref{linear_resp}. 

\section{Determination of subspace decomposition parameter $A$}\label{append:A_opt}

The parameter $A$ which defines our subsystem DFT functional (Eq.~(\ref{def_NN_sub})) can be 
determined by minimizing the cost function $L$ (Eq.~(\ref{cost_function})). In the case, the metric tensor $G_{ik}$ (Eq.~(\ref{def_metric})) is expanded to $(n_W + 1) \times (n_W + 1) $ dimension by adding the components related to $\partial [{\delta \tilde{T}^{\rm{NN}}({\bm r}_{p})}/{\delta\rho}]/\partial A$. We have indeed minimized the cost function with the expanded metric tensor and determined the parameter $A$. 
In the actual minimization, we have performed grid search for the initial values of $A_m = 10^{0.5 m}$ with $m$ being integers from 0 to 9 (a typical logarithmic grid) and the subsequent neural-network minimization of the cost function $L$. 
We have found that the optimized value $A^{\rm opt}$ depends on the initial choice $A_m$ as shown in Table \ref{tab_A_interp}, 
indicating that the cost function $L$ shows multi-stability as a function of $A$. 
This is presumably due to a fact that the NN weights $\{ W^i \}$ rather than $A$ are decisive to determine the cost function. The computed RMSE of our KEFD with respect to the KS-KEFD, i.e., $\sqrt{2L}$ 
itself, for each value of $A^{\rm opt}$ are tabulated in Table \ref{tab_A_interp}. For $A$ = 10.549, $\sqrt{2L}$ has the minimum value of 0.269 Ha although it increases only by 7\% for $A$ = 31.76.

\begin{table}[!h]
\begin{center}
\caption{The subsystem-DFT parameter $A$ and the corresponding RMSE between our NN KEFD and KS KEFD (the column ``$\sqrt{2L}$'' in the unit of Ha). KS density $\rho^{\rm KS}$ is used to evaluate $\sqrt{2L}$. The RMSE of the SCF density $\rho_{\rm SCF}$ obtained by our OFDFT scheme with each value of $A$ with respect to $\rho^{\rm KS}$ is also shown (the column ``RMSE'' in the unit of $10^{-2} \times {\rm bohr}^{-3}$). $[ (n_W + 1) \times (n_W + 1) ] {\bm G}$ and [$n_W \times n_W] {\bm G}$ denote the results with the expanded metric tensor and the original metric tensor without the $\partial [{\delta \tilde{T}^{\rm{NN}}({\bm r}_{p})}/{\delta\rho}]/\partial A$-related components, respectively.}
\begin{tabular}{c|ccc|cc} \hline\hline
$m$ &  \multicolumn{3}{c|}{$[ (n_W + 1) \times (n_W + 1) ] {\bm G}$} & \multicolumn{2}{c}{[$n_W \times n_W] {\bm G}$}\\
($A_m = 10^{0.5 m}$) &  \ $A^{\rm opt}$ & \ \ \ $\sqrt{2L}$ \ \ & 
RMSE  & \ \ \ $\sqrt{2L}$ \ \ & RMSE
\\ \hline
 0 & 2.751020                 & 0.433 & 1.904 & 0.289 & 1.271   \\ 
 1 & 4.220258                   & 0.331 & 1.534 & 0.273 & 1.262   \\ 
 2 & 10.54878    & 0.269 & 1.225 & 0.264 & 1.201   \\ 
 3 & 31.76219    & 0.284 & 1.232 & 0.274 & 1.186   \\ 
 4 & 100.0558  & 0.328 & 1.204 & 0.326 & 1.194   \\ 
 5 & 316.2413  & 0.404 & 1.300 & 0.387 & 1.246   \\ 
 6 & 1000.005    & 0.408 & 1.274 & 0.400 & 1.246   \\ 
 7 & 3162.279    & 0.337 & 1.231 & 0.346 & 1.266    \\ 
 8 & 10000.00  & 0.297 & 1.243 & 0.297 & 1.243    \\ 
 9 & 31622.78  & 0.279 & 1.238 & 0.279 & 1.238    \\ 
\hline\hline
\end{tabular}
\label{tab_A_interp}
\end{center}
\end{table}

Observing the insensitivity of the cost function to the parameter $A$, we 
have also performed the minimization of the cost function with the original metric tensor with the dimension of $n_W$ 
with the fixed value of $A = A_m$. The obtained 
$\sqrt{2L}$ is shown in Table \ref{tab_A_interp}. The $\sqrt{2L}$ values are comparable with those obtained for $A^{\rm opt}$ and the minimum value 0.264 Ha for $A_m$ = 10 is even smaller than the value 0.269 Ha for $A$ = 10.549 above. This is presumably 
because we have succeeded to optimize $\{ W^i \}$ better 
with the fixed choice of $A$. This may indicate the limitation of the present SNGD optimization scheme for the 
two parameter sets with different mathematical structures. 

Having observed that the $\sqrt{2L}$ values are insensitive 
to the value of $A$, we introduce another criterion to determine the $A$ value. That is to minimize the difference of the electron density $\rho_{\rm SCF}$ obtained by the SCF calculations in our OFDFT scheme with each $A$ and $\{ W^i \}$ values from that $\rho^{\rm KS}$ obtained by the KSDFT scheme. The minimum RMSE 
1.186 $\times 10^{-2}$ bohr$^{-3}$ 
of those two densities are obtained for the $A$ value of 10$^{1.5}$. 
This set of parameters, $A$ and $\{ W^i \}$ in turn produces value of 
0.274 Ha for $\sqrt{2L}$ which is just 3.8 \% larger than the minimum $\sqrt{2L}$ 
explained above. 
Recalling that the electron density is the fundamental quantity in DFT, we adopt the value of $A$ = 10$^{1.5}$ in 
the present work. 
Further examination of the $A$ value remains in future. 

\clearpage
\bibliography{main_arxiv}

\clearpage
\pagebreak
\title{ Supplemental Material for ``Order-$N$ orbital-free density-functional calculations with machine learning of functional derivatives for semiconductors and metals"}
\maketitle

\setcounter{secnumdepth}{1}
\renewcommand{\thesection}{\Roman{section}}
\renewcommand{\theequation}{S\arabic{equation}}
\renewcommand{\thefigure}{S\arabic{figure}}
\renewcommand{\thetable}{S\arabic{table}}
\setcounter{figure}{0}

\onecolumngrid
In this supplemental material, we show the calculated Self-Consistent-Field (SCF) electron densities obtained by our NN KEDF along with those obtained by using the KEDFs in the past, i.e., PGSL0.25, LKT, and the conventional TF(1/5)vW [TF($\lambda$)vW with $\lambda$ = 1/5] KEDFs. As is demonstrated by RMSE of the SCF densities with respect to the KS density in Tables II, III, and IV in the main text, our NN KEDFs outperform the previous KEDFs. The SCF density of diamond-structured (ds-) Si shown in Fig. 5 in the main text also shows the superiority of the present NN KEDF. 
Here Figs. S1-S23 represent the SCF densities of other 23 systems which corroborate the superiority of the NN KEDF: The densities of ds-C, graphene, fcc-Si, $\beta$-tin Si, zincblende(3C)-SiC, bcc-Li, fcc-Al, fcc-Cu, bcc-Na, NaCl, Li2, C2, Na2, Al2, Si2, Cl2, Li atom, C atom, Na atom, Al atom, Si atom, Cl atom, and Cu atom. 

\begin{figure}[!h]
\centering
\includegraphics[width=.85\linewidth]{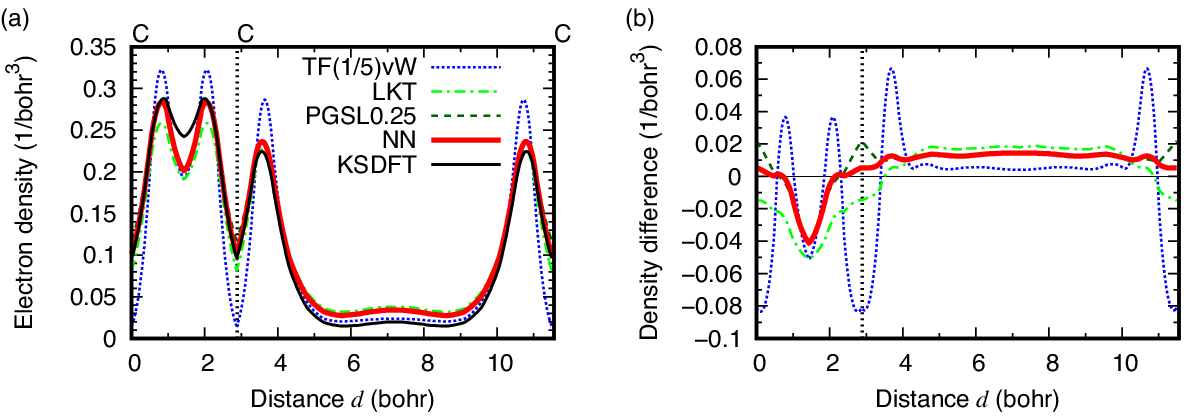}
\caption{
(a) SCF electron density in diamond along [111] direction obtained by different approximations to KEDF. 
The horizontal axis is the distance $d$ from a C atom along [111]. C atoms are located at the positions depicted by the vertical dashed line, and the left and right ends. 
(b) Difference of densities with respect to the KS density (black solid line in (a)). The 0 value on the ordinate is indicated by a horizontal black solid line.
}\label{rho_c2_dia_compare}
\end{figure}
%
\begin{figure}[!h]
\centering
\includegraphics[width=.85\linewidth]{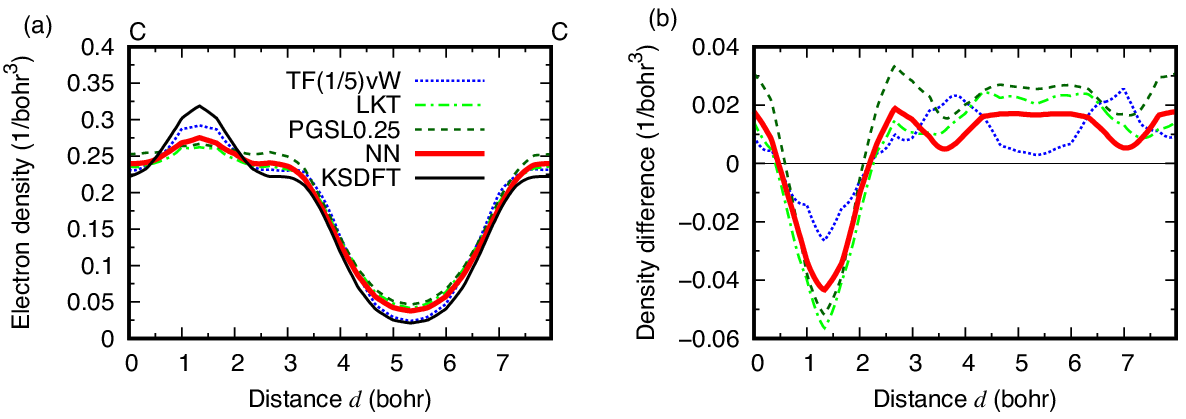}
\caption{
(a) SCF electron density in graphene along C-C bond direction obtained by different approximations to KEDF. 
The horizontal axis is the distance $d$ from a C atom. C atoms are located at the left and right ends. 
(b) Difference of densities with respect to the KS density (black solid line in (a)). The 0 value on the ordinate is indicated by a horizontal black solid line.
}\label{rho_graphene_compare}
\end{figure}
%
\begin{figure}[!h]
\centering
\includegraphics[width=.85\linewidth]{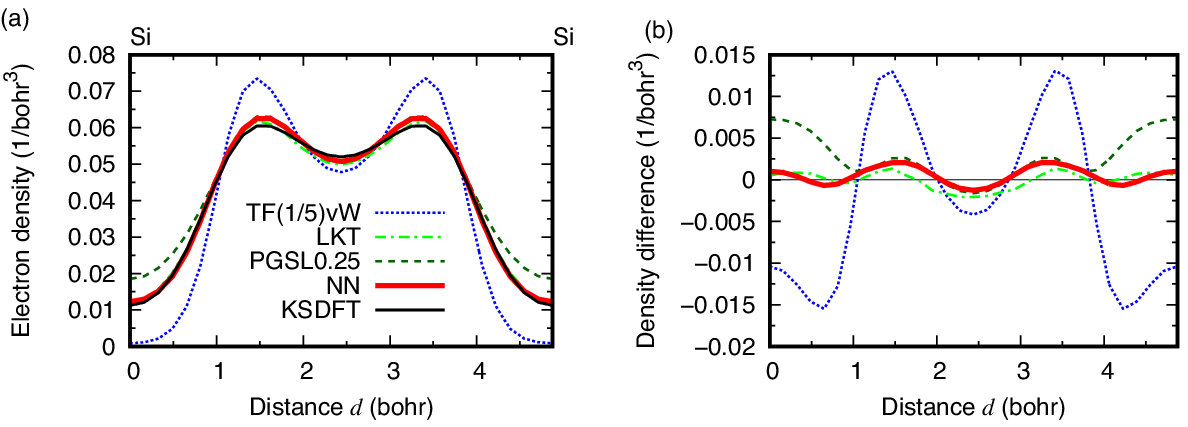}
\caption{
(a) SCF electron density in fcc-Si along Si-Si bond direction obtained by different approximations to KEDF. 
The horizontal axis is the distance $d$ from a Si atom along the bond direction. Si atoms are located at the left and right ends. 
(b) Difference of densities with respect to the KS density (black solid line in (a)). The 0 value on the ordinate is indicated by a horizontal black solid line.
}\label{rho_fcc-si_compare}
\end{figure}
%
\begin{figure}[!h]
\centering
\includegraphics[width=.85\linewidth]{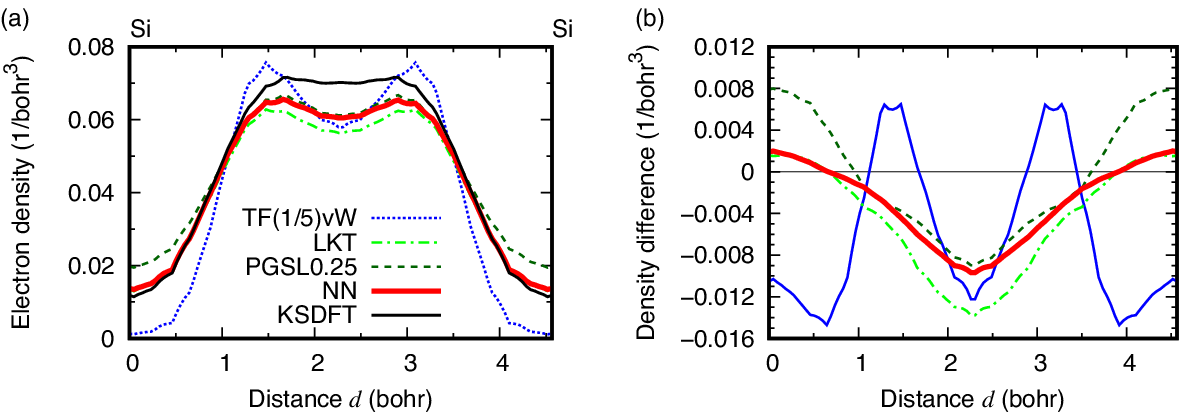}
\caption{
(a) SCF electron density in $\beta$-tin Si along Si-Si bond direction obtained by different approximations to KEDF. 
The horizontal axis is the distance $d$ from a Si atom along the bond direction. Si atoms are located at the left and right ends. 
(b) Difference of densities with respect to the KS density (black solid line in (a)). The 0 value on the ordinate is indicated by a horizontal black solid line.
}\label{rho_b-si_dia_compare}
\end{figure}
%
\begin{figure}[!h]
\centering
\includegraphics[width=.85\linewidth]{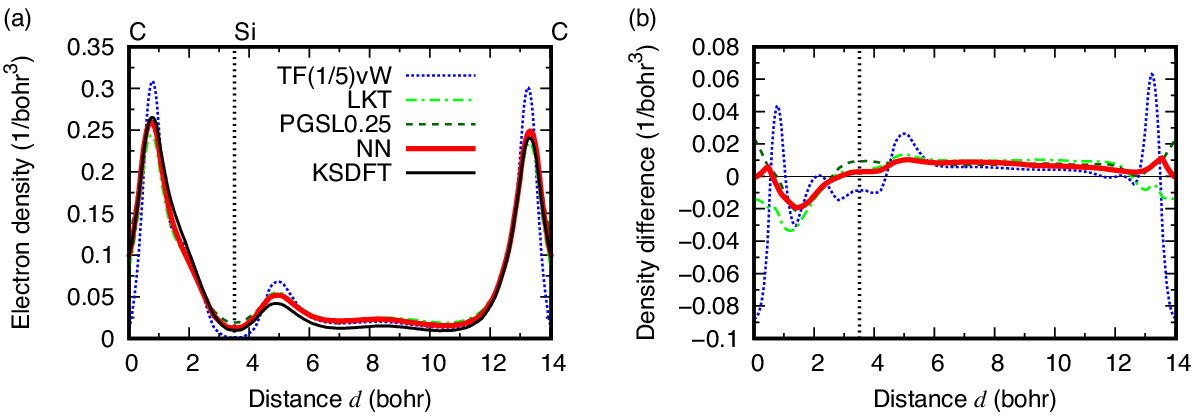}
\caption{
(a) SCF electron density in 3C-SiC along [111] direction obtained by different approximations to KEDF. 
The horizontal axis is the distance $d$ from a C atom along [111]. C atoms are located at the left and right ends, whereas a Si atom is located at the position depicted by the vertical dashed line. 
(b) Difference of densities with respect to the KS density (black solid line in (a)). The 0 value on the ordinate is indicated by a horizontal black solid line.
}\label{rho_sic_compare}
\end{figure}
%
\begin{figure}[!h]
\centering
\includegraphics[width=.85\linewidth]{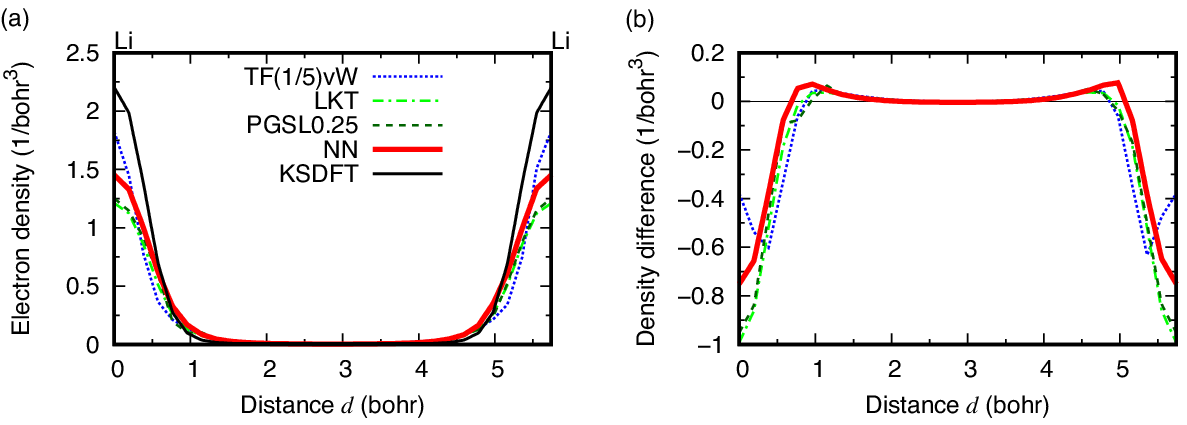}
\caption{
(a) SCF electron density in bcc-Li along [100] direction obtained by different approximations to KEDF. 
The horizontal axis is the distance $d$ from a Li atom along [100]. Li atoms are located at the left and right ends. 
(b) Difference of densities with respect to the KS density (black solid line in (a)). The 0 value on the ordinate is indicated by a horizontal black solid line.
}\label{rho_li_compare}
\end{figure}
%
\begin{figure}[!h]
\centering
\includegraphics[width=.85\linewidth]{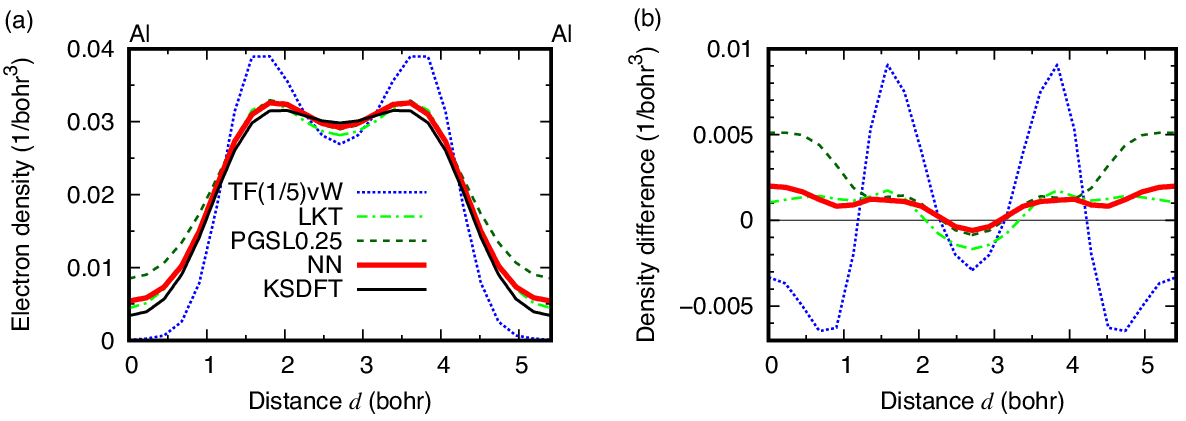}
\caption{
(a) SCF electron density in fcc-Al along [100] direction obtained by different approximations to KEDF. 
The horizontal axis is the distance $d$ from an Al atom along [100]. Al atoms are located at the left and right ends. 
(b) Difference of densities with respect to the KS density (black solid line in (a)). The 0 value on the ordinate is indicated by a horizontal black solid line.
}\label{rho_al_compare}
\end{figure}
%
\begin{figure}[!h]
\centering
\includegraphics[width=.85\linewidth]{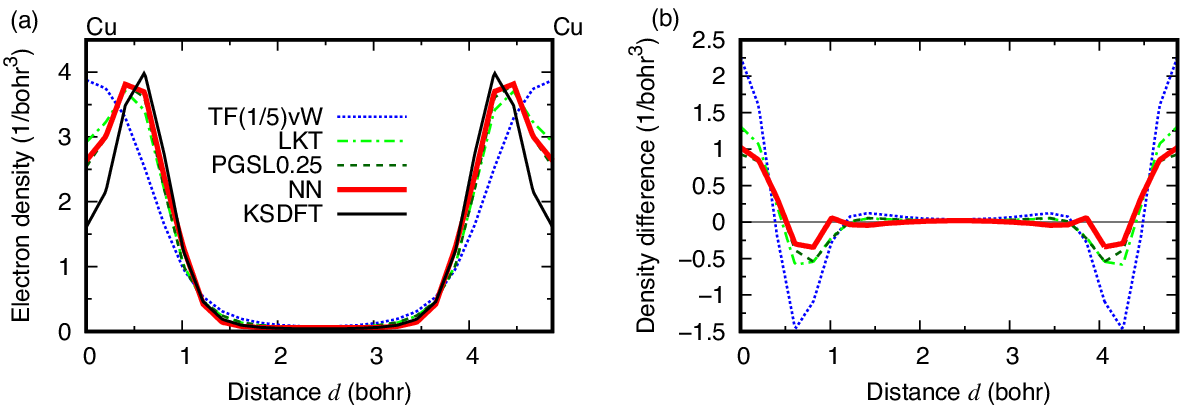}
\caption{
(a) SCF electron density in fcc-Cu along [100] direction obtained by different approximations to KEDF. 
The horizontal axis is the distance $d$ from a Cu atom along [100]. Cu atoms are located at the left and right ends. 
(b) Difference of densities with respect to the KS density (black solid line in (a)). The 0 value on the ordinate is indicated by a horizontal black solid line.
}\label{rho_cu_compare}
\end{figure}
%
\begin{figure}[!h]
\centering
\includegraphics[width=.85\linewidth]{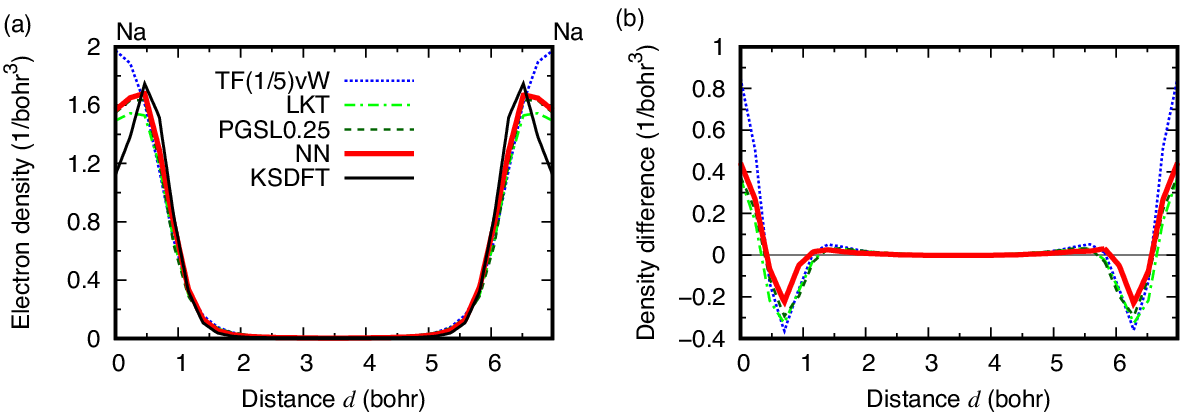}
\caption{
(a) SCF electron density in bcc-Na along [100] direction obtained by different approximations to KEDF. 
The horizontal axis is the distance $d$ from a Na atom along [100]. Na atoms are located at the left and right ends. 
(b) Difference of densities with respect to the KS density (black solid line in (a)). The 0 value on the ordinate is indicated by a horizontal black solid line.
}\label{rho_na_compare}
\end{figure}
%
\begin{figure}[!h]
\centering
\includegraphics[width=.85\linewidth]{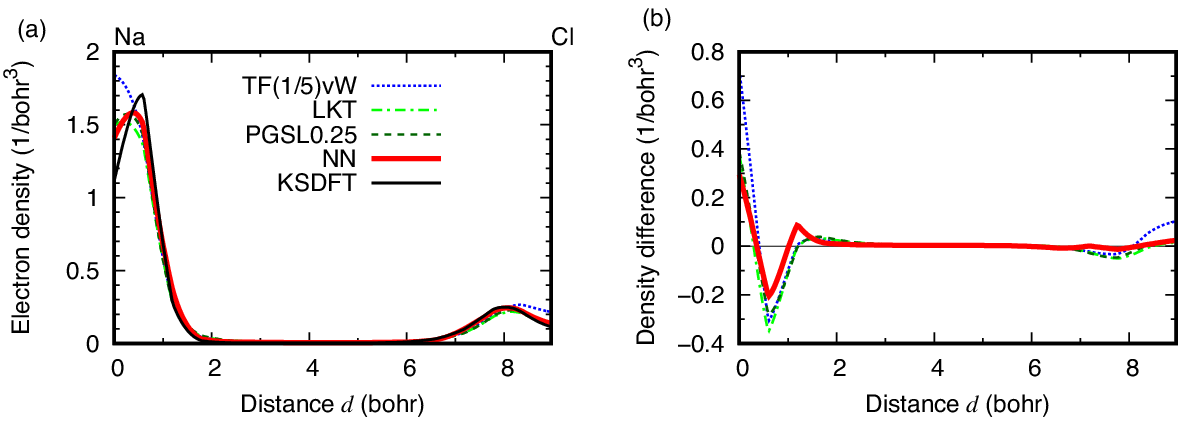}
\caption{
(a) SCF electron density in NaCl along [100] direction obtained by different approximations to KEDF. 
The horizontal axis is the distance $d$ from a Na atom along [100].
(b) Difference of densities with respect to the KS density (black solid line in (a)). The 0 value on the ordinate is indicated by a horizontal black solid line.
}\label{rho_nacl_compare}
\end{figure}
%
\begin{figure}[!h]
\centering
\includegraphics[width=.85\linewidth]{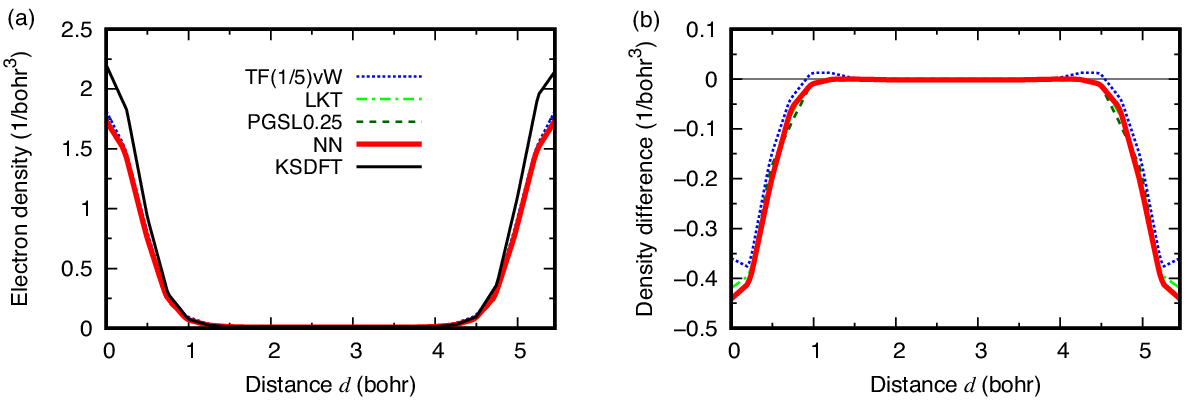}
\caption{
(a) SCF electron density in Li$_2$ along Li-Li bond direction obtained by different approximations to KEDF. 
The horizontal axis is the distance $d$ from the Li nucleus. Li atoms are located at the left and right end, respectively. 
(b) Difference of densities with respect to the KS density (black solid line in (a)). The 0 value on the ordinate is indicated by a horizontal black solid line.
}\label{rho_li-li_compare}
\end{figure}
%
\begin{figure}[!h]
\centering
\includegraphics[width=.85\linewidth]{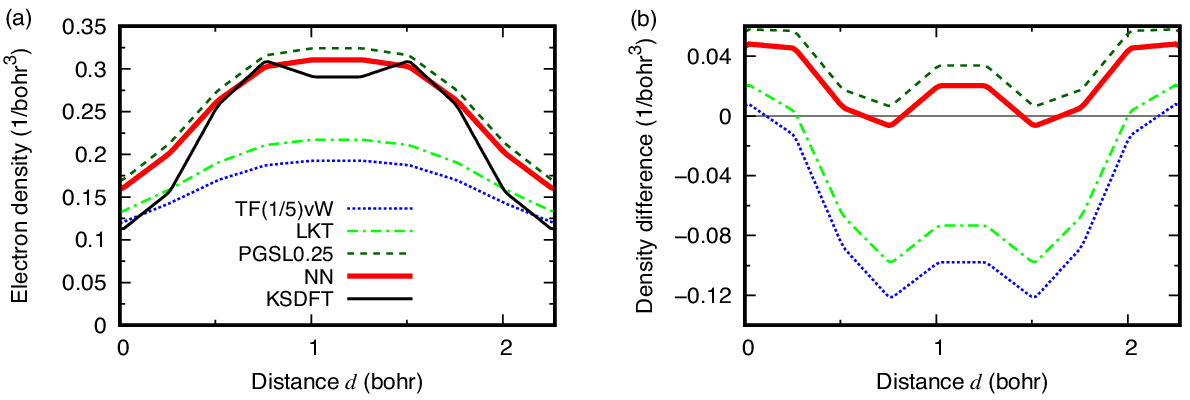}
\caption{
(a) SCF electron density in C$_2$ along C-C bond direction obtained by different approximations to KEDF. 
The horizontal axis is the distance $d$ from the C nucleus. C atoms are located at the left and right end, respectively. 
(b) Difference of densities with respect to the KS density (black solid line in (a)). The 0 value on the ordinate is indicated by a horizontal black solid line.
}\label{rho_c-c_compare}
\end{figure}
%
\begin{figure}[!h]
\centering
\includegraphics[width=.85\linewidth]{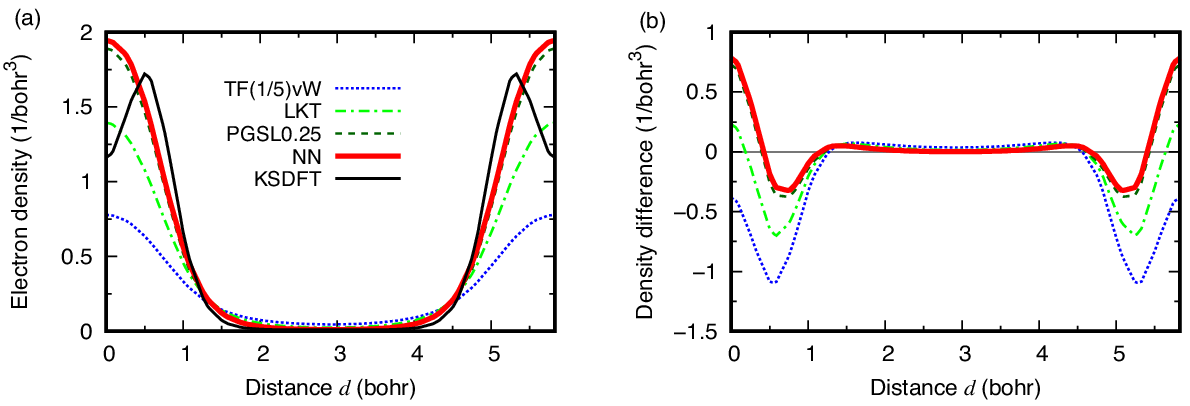}
\caption{
(a) SCF electron density in Na$_2$ along Na-Na bond direction obtained by different approximations to KEDF. 
The horizontal axis is the distance $d$ from the Na nucleus. Na atoms are located at the left and right end, respectively. 
(b) Difference of densities with respect to the KS density (black solid line in (a)). The 0 value on the ordinate is indicated by a horizontal black solid line.
}\label{rho_na-na_compare}
\end{figure}
\clearpage
\begin{figure}[!h]
\centering
\includegraphics[width=.85\linewidth]{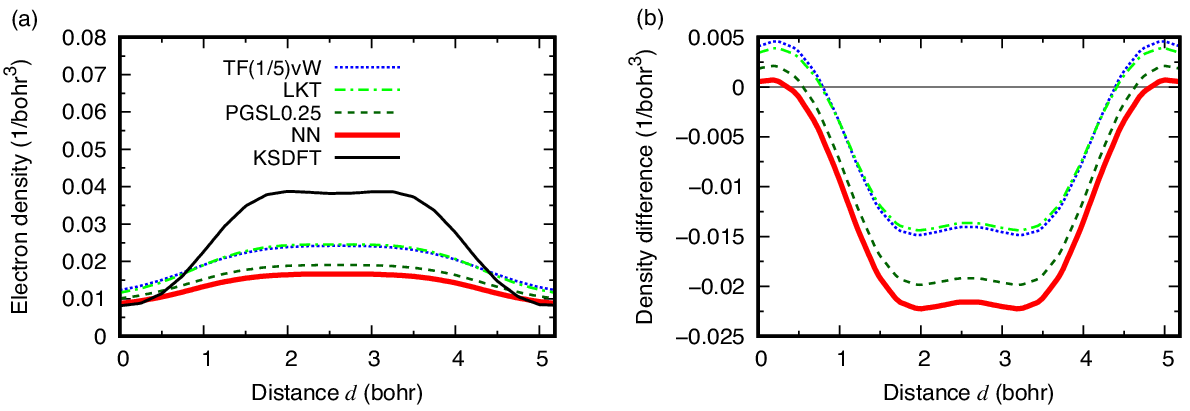}
\caption{
(a) SCF electron density in Al$_2$ along Al-Al bond direction obtained by different approximations to KEDF. 
The horizontal axis is the distance $d$ from the Al nucleus. Al atoms are located at the left and right end, respectively. 
(b) Difference of densities with respect to the KS density (black solid line in (a)). The 0 value on the ordinate is indicated by a horizontal black solid line.
}\label{rho_al-al_compare}
\end{figure}
%
\begin{figure}[!h]
\centering
\includegraphics[width=.85\linewidth]{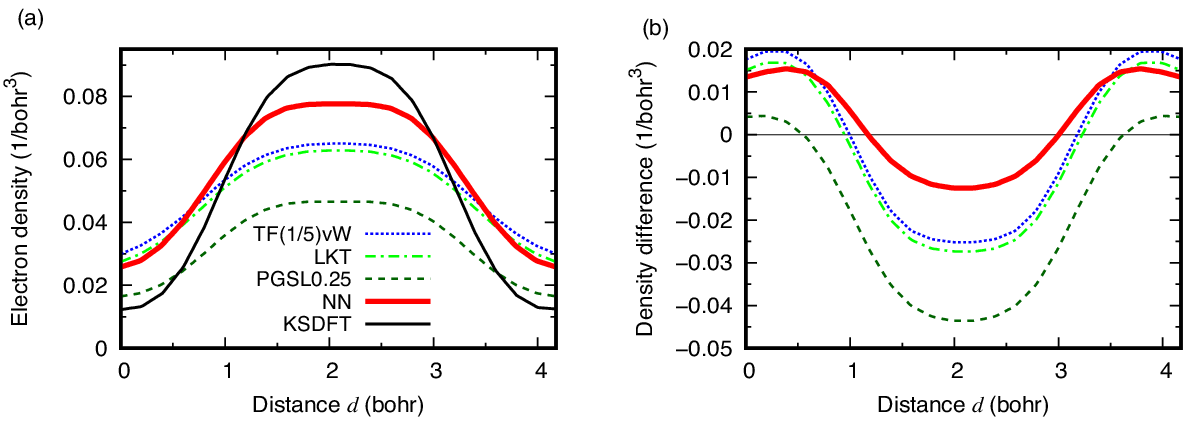}
\caption{
(a) SCF electron density in Si$_2$ along Si-Si bond direction obtained by different approximations to KEDF. 
The horizontal axis is the distance $d$ from the Si nucleus. Si atoms are located at the left and right end, respectively. 
(b) Difference of densities with respect to the KS density (black solid line in (a)). The 0 value on the ordinate is indicated by a horizontal black solid line.
}\label{rho_si-si_compare}
\end{figure}
\clearpage
\begin{figure}[!h]
\centering
\includegraphics[width=.85\linewidth]{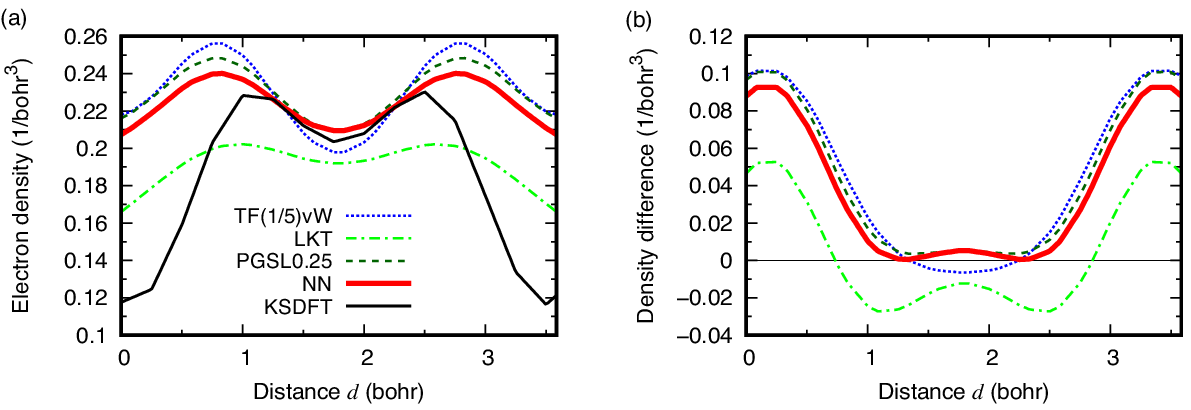}
\caption{
(a) SCF electron density in Cl$_2$ along Cl-Cl bond direction obtained by different approximations to KEDF. 
The horizontal axis is the distance $d$ from the Cl nucleus. Cl atoms are located at the left and right end, respectively. 
(b) Difference of densities with respect to the KS density (black solid line in (a)). The 0 value on the ordinate is indicated by a horizontal black solid line.
}\label{rho_cl-cl_compare}
\end{figure}
\FloatBarrier
\begin{figure}[!h]
\centering
\includegraphics[width=.85\linewidth]{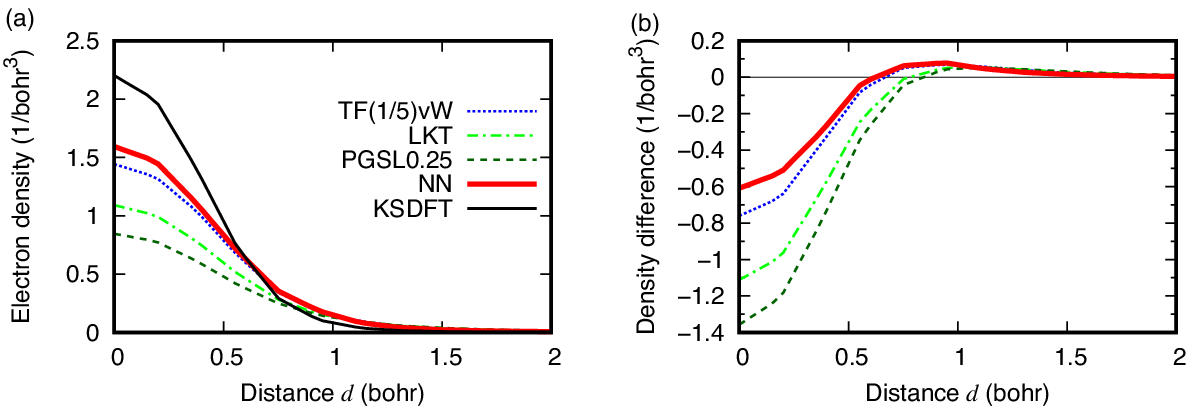}
\caption{
(a) Radial SCF electron density in Li atom obtained by different approximations to KEDF. 
The horizontal axis is the distance $d$ from the nucleus.
(b) Difference of densities with respect to the KS density (black solid line in (a)). The 0 value on the ordinate is indicated by a horizontal black solid line.
}\label{rho_li_atom_compare}
\end{figure}
\clearpage
\begin{figure}[!h]
\centering
\includegraphics[width=.85\linewidth]{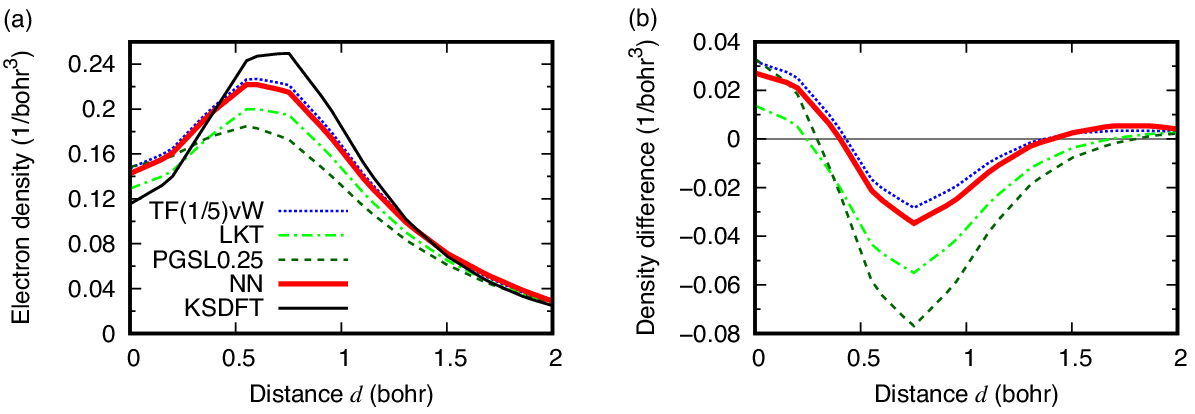}
\caption{
(a) Radial SCF electron density in C atom obtained by different approximations to KEDF. 
The horizontal axis is the distance $d$ from the nucleus.
(b) Difference of densities with respect to the KS density (black solid line in (a)). The 0 value on the ordinate is indicated by a horizontal black solid line.
}\label{rho_c_atom_compare}
\end{figure}

\begin{figure}[!h]
\centering
\includegraphics[width=.85\linewidth]{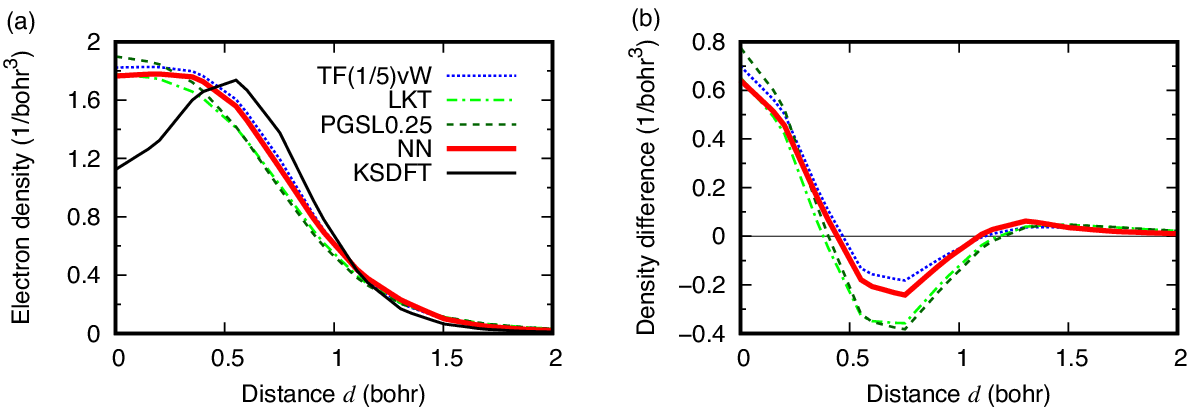}
\caption{
(a) Radial SCF electron density in Na atom obtained by different approximations to KEDF. 
The horizontal axis is the distance $d$ from the nucleus.
(b) Difference of densities with respect to the KS density (black solid line in (a)). The 0 value on the ordinate is indicated by a horizontal black solid line.
}\label{rho_na_atom_compare}
\end{figure}
%
\begin{figure}[!h]
\centering
\includegraphics[width=.85\linewidth]{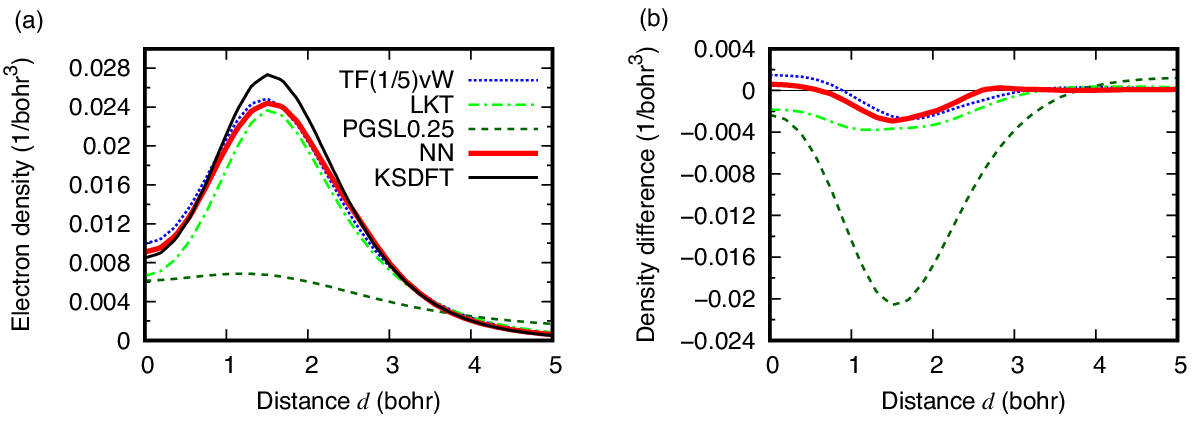}
\caption{
(a) Radial SCF electron density in Al atom obtained by different approximations to KEDF. 
The horizontal axis is the distance $d$ from the nucleus.
(b) Difference of densities with respect to the KS density (black solid line in (a)). The 0 value on the ordinate is indicated by a horizontal black solid line.
}\label{rho_al_atom_compare}
\end{figure}
%
\begin{figure}[!h]
\centering
\includegraphics[width=.85\linewidth]{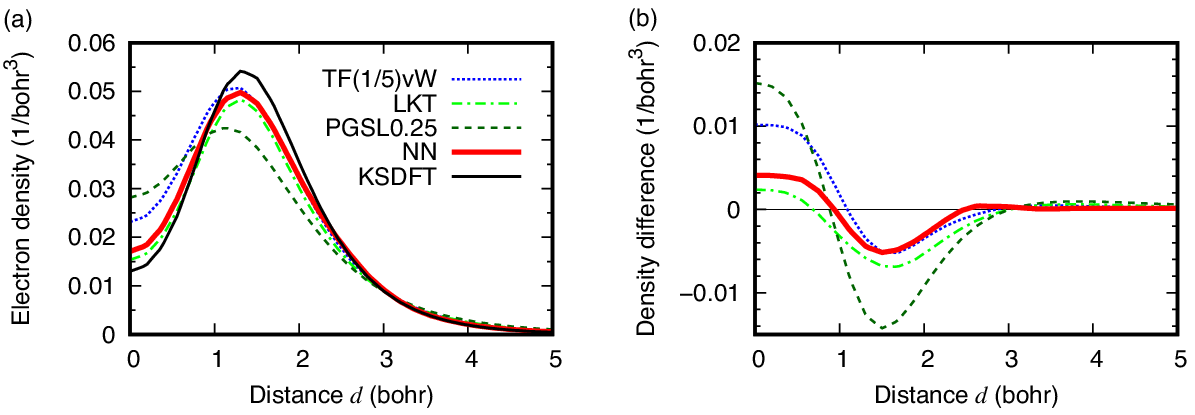}
\caption{
(a) Radial SCF electron density in Si atom obtained by different approximations to KEDF. 
The horizontal axis is the distance $d$ from the nucleus.
(b) Difference of densities with respect to the KS density (black solid line in (a)). The 0 value on the ordinate is indicated by a horizontal black solid line.
}\label{rho_si_atom_compare}
\end{figure}
\FloatBarrier
%
\begin{figure}[!h]
\centering
\includegraphics[width=.85\linewidth]{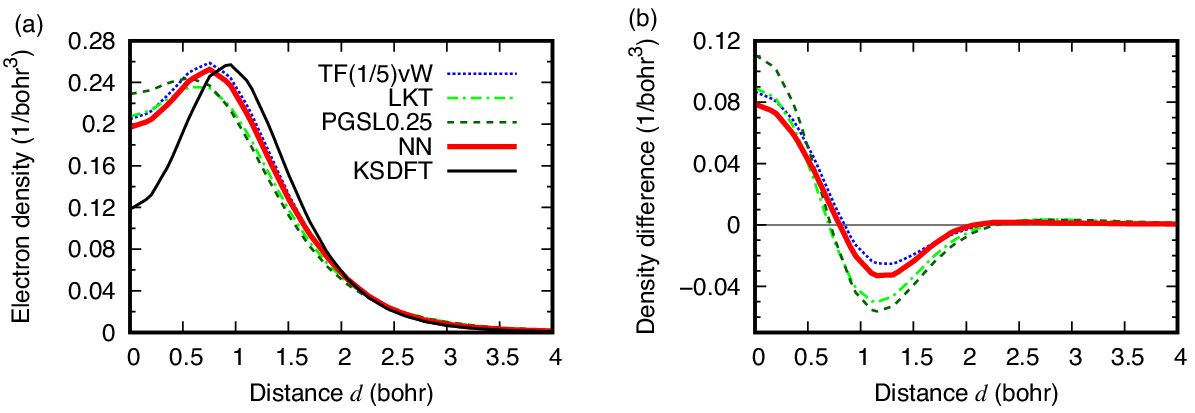}
\caption{
(a) Radial SCF electron density in Cl atom obtained by different approximations to KEDF. 
The horizontal axis is the distance $d$ from the nucleus.
(b) Difference of densities with respect to the KS density (black solid line in (a)). The 0 value on the ordinate is indicated by a horizontal black solid line.
}\label{rho_cl_atom_compare}
\end{figure}
%
\begin{figure}[!h]
\centering
\includegraphics[width=.85\linewidth]{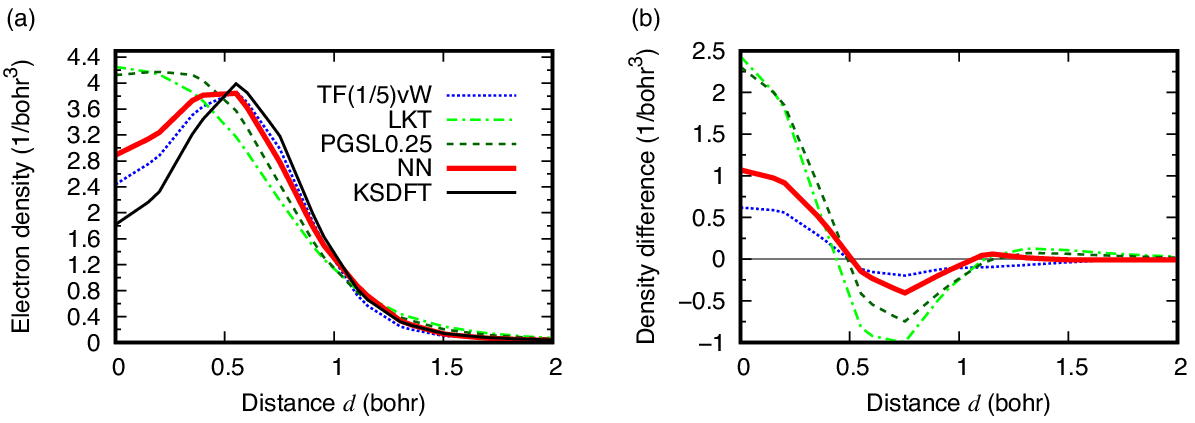}
\caption{
(a) Radial SCF electron density in Cu atom obtained by different approximations to KEDF. 
The horizontal axis is the distance $d$ from the nucleus.
(b) Difference of densities with respect to the KS density (black solid line in (a)). The 0 value on the ordinate is indicated by a horizontal black solid line.
}\label{rho_cu_atom_compare}
\end{figure}
\FloatBarrier
\clearpage

\end{document}